\documentclass[journal]{IEEEtran}
\usepackage{amsmath,amsfonts}
\usepackage{tablefootnote}
\usepackage{threeparttable}
\usepackage{amssymb}
\usepackage{algorithmic}
\usepackage{algorithm}
\usepackage{array}
\usepackage[caption=false,font=normalsize,labelfont=sf,textfont=sf]{subfig}
\usepackage[colorlinks,linkcolor=red,]{hyperref}
\usepackage{textcomp}
\usepackage{stfloats}
\usepackage{lineno}
\usepackage{url}
\usepackage{caption}
\usepackage{verbatim}
\usepackage{graphicx}
\usepackage{cite}
\usepackage{bm}
\usepackage{xcolor}
\usepackage{multirow}
\usepackage{booktabs}
\usepackage{enumerate}
\usepackage{siunitx}
\usepackage{gensymb}
\begin{document}

\title{Fish Tracking, Counting, and Behaviour Analysis in Digital Aquaculture: A Comprehensive Survey}

\author{Meng Cui$^{1}$, Xubo Liu$^{1}$, Haohe Liu$^{1}$, Jinzheng Zhao$^{1}$,  
Daoliang Li$^{2}$, Wenwu Wang$^{1}$
        % <-this % stops a space
\thanks{M. Cui, X.Liu, H. Liu, J. Zhao, and W. Wang are with the Centre for Vision, Speech and Signal Processing (CVSSP), University of Surrey, Guildford GU2 7XH, UK. (e-mail:
[m.cui, xubo.liu, haohe.liu, j.zhao, w.wang]@surrey.ac.uk).}
% \thanks{T. Chen and G. Lian are with the Department of Chemical and Process Engineering, University of Surrey, Guilford GU2 7XH, UK. (e-mail: [t.chen, g.lian]@surrey.ac.uk).}
\thanks{D. Li are with the National Innovation Center for Digital Fishery, China Agricultural University, China (e-mail: dliangl@cau.edu.cn).}}

% \pagewiselinenumbers % 按页重新编号 
% \switchlinenumbers

\maketitle

\begin{abstract}
Digital aquaculture leverages advanced technologies and data-driven methods, providing substantial benefits over traditional aquaculture practices. \textcolor{black}{This paper presents a comprehensive review of three interconnected digital aquaculture tasks, namely, fish tracking, counting, and behaviour analysis, using a novel and unified approach. Unlike previous reviews which focused on single modalities or individual tasks, we analyse vision-based (i.e. image- and video-based), acoustic-based, and biosensor-based methods across all three tasks. We examine their advantages, limitations, and applications, highlighting recent advancements and identifying critical cross-cutting research gaps. The review also includes emerging ideas such as applying multi-task learning and large language models to address various aspects of fish monitoring, an approach not previously explored in aquaculture literature. We identify the major obstacles hindering research progress in this field, including the scarcity of comprehensive fish datasets and the lack of unified evaluation standards. To overcome the current limitations, we explore the potential of using emerging technologies such as multimodal data fusion and deep learning to improve the accuracy, robustness, and efficiency of integrated fish monitoring systems. In addition, we provide a summary of existing datasets available for fish tracking, counting, and behaviour analysis. This holistic perspective offers a roadmap for future research, emphasizing the need for comprehensive datasets and evaluation standards to facilitate meaningful comparisons between technologies and to promote their practical implementations in real-world settings.}

\end{abstract}


\section{Introduction}
\label{sec:intro}

With the expansion of the global population and the degradation of the ecological environment, traditional fishing (i.e. capture fisheries) is no longer capable of meeting the growing human demand for fish products \cite{clavelle2019interactions, tacon2020trends}. Aquaculture has become the primary source of fish acquisition, and digital aquaculture is emerging as a promising approach to enhance the efficiency and sustainability of the industry \cite{li2024recent}. 

\textcolor{black}{As enabling technologies in digital aquaculture, fish tracking, counting, and behaviour analysis have been studied extensively.} Traditional methods for fish tracking and behaviour analysis rely on the experience of human observers, and the observation results depend on their skills and knowledge, which are not always reliable \cite{an2021survey, duarte2009measurement}.
 Similarly, manual fish counting methods involve removing fish from tanks, leading to stress, injury, and disease, negatively impacting fish welfare and growth \cite{zhang2020automatic, zhou2018intelligent}. The implementation of intelligent tracking, counting, and behaviour analysis technologies can help overcome these limitations, reducing the risk of fish mortality, improving feeding strategies, and promoting sustainable development in aquaculture \cite{politikos2021movclufish, li2021automatic, puig2019automatic}. \textcolor{black}{This paper aims to provide a comprehensive survey of the research problems, existing technologies, their connections, and potential gaps in this field. }
 
\textcolor{black}{There are already literature reviews conducted previously in this field, such as \cite{delcourt2013video, xia2018aquatic, li2021automatic, yang2021computer, li2022recent, mei2022recent}. However, most of them focused on individual tasks or single modalities, primarily emphasizing computer vision technology as the main approach. In practice, these tasks are interconnected and form a coherent part of an overall digital aquaculture system. For example, accurate monitoring of these tasks is vital for detecting abnormal fish behaviour, estimating fish abundance, and formulating effective management strategies, ultimately improving fish welfare and economic outcomes \cite{cui2022fish}. The narrow focus in previous reviews is inherently limited in capturing the interconnected nature of these tasks in real-world aquaculture scenarios. Different from these reviews, our focus is not only on surveying the progress made for these tasks, but also on establishing the connections among them.}

 Various technologies such as vision-based, acoustic-based and biosensing-based methods are used for fish tracking, counting, and behaviour analysis in aquaculture. Vision-based sensors and computer vision technology have found widespread application due to advancements in optical imaging and computer vision. However, they are limited by poor illumination, low contrast, high noise, fish deformation, frequent occlusion, and dynamic backgrounds \cite{li2022recent, zhang2020estimation, soltanzadeh2020prototype, yang2021deep}. Acoustic-based sensors and hydroacoustic methods, which are non-invasive, are beneficial for monitoring fish in turbid water environments and overnight, but their high hardware cost limits their popularity in intensive aquaculture settings \cite{helminen2021object, capoccioni2019fish, eggleston2020improved, colborne2019sequence, zhou2018near}. Biosensors can provide valuable information on fish physiology and behaviour, but their invasive nature and the need for individual fish tagging can be challenging in large-scale aquaculture operations and may potentially harm the fish \cite{kolarevic2021novel}.

\textcolor{black}{A systematic review of literature from 2000 to 2023 reveals significant growth in research related to these technologies. Annual publications have increased from around 200 in the early 2000s to over 2,000 in recent years. Computer vision techniques dominate the field, accounting for about 40\% of the publications, followed by acoustic monitoring methods and behavioural analysis approaches, each representing about a quarter of the research outputs. The geographical distribution of research and methodological preferences show interesting patterns. Asia, particularly China and Japan, leads in publications, contributing nearly half of the global research outputs, mostly on computer vision techniques in tracking and counting, driven by the prevalence of large-scale intensive aquaculture operations and robust technological infrastructure. Europe, contributing about 30\% of publications, emerges as a leader in acoustic monitoring and behavioural analysis techniques. This trend aligns with the region's stringent welfare regulations and emphasis on environmental sustainability in aquaculture practices. North America, while producing fewer publications, pioneers the integration of machine learning with various monitoring methods, leveraging its strong artificial intelligence (AI) research base and focus on automation. However, this global research landscape also reveals significant gaps. Developing regions, including parts of Africa, Asia, and Latin America, show limited adoption of advanced monitoring technologies. This gap can be attributed to the high initial costs, the lack of technical expertise, and the predominance of small-scale farming in these areas. Oceania, despite its significant aquaculture industry, is underrepresented in research outputs, possibly due to a smaller research community and a focus on practical applications over academic publications. Eastern Europe and Central Asia also show fewer publications on novel monitoring technologies, likely due to the later development of intensive aquaculture in these regions and limited research funding.}

 To address these limitations, our paper takes a holistic approach by systematically surveying a range of technologies, including vision-based sensors, acoustic-based sensors, biosensors, and hydroacoustic methods. We consider fish tracking, counting, and behaviour analysis as interconnected components of a unified task in digital aquaculture. This comprehensive perspective facilitates a more integrated discussion of these tasks while identifying technology gaps in the current literature.
Furthermore, we explore the potential of emerging technologies such as multimodal data fusion, deep learning, and multi-task learning models that can simultaneously handle multiple aspects of fish monitoring. We also discuss the integration of large language models (LLM) and their potential to enhance understanding and decision-making in digital aquaculture. By adopting this multifaceted approach, our review aims to provide a more complete understanding of the challenges and opportunities in digital aquaculture, paving the way for developing more efficient, accurate, and integrated fish monitoring systems. The insights provided in this review have the potential to guide future research directions, inform policy decisions, and ultimately contribute to the sustainable growth of the aquaculture industry.

% This article comprehensively reviews the literature on fish tracking, counting, and behavioural analysis in aquaculture over the past two decades, emphasizing the progress made in these areas and identifying potential future research directions. 
The remainder of this article is structured as follows: Section \ref{sec: tracking} explores the advancements in fish tracking techniques, while Section \ref{sec: counting} discusses the various methods and applications of fish counting. Section \ref{sec: behaviour} discusses the behaviour analysis of fish, and Section \ref{sec:multimodal} introduces the potential of multimodal fusion in aquaculture. Section \ref{sec: dataset} presents an overview of relevant public datasets. In Section \ref{sec: challenges}, we examine the challenges faced by the aquaculture industry and discuss future development trends. Finally, Section \ref{sec: conclusion} summarises the key findings and conclusions presented in this paper.

\section{Fish tracking}
\label{sec: tracking}
Vision-based multi-target tracking methods are increasingly used in fish behaviour analysis. However, fish tracking is challenging because of the small differences between individuals, complex environments, and variations in plankton, shapes, angles, and scales of swimming fish \cite{shreesha2020computer}. Fish tracking can be categorized into two-dimensional (2D) and three-dimensional (3D) tracking based on the swimming environment \cite{qian2017feature}. 2D tracking is used in shallow water containers, where fish swimming appears close to a 2D planar motion and is represented using (x, y) coordinates. Still, it can only analyse a part of the fish's behaviour. In contrast, 3D tracking considers depth information and is represented using (x, y, z) coordinates, enabling the analysis of spatial movement in natural environments.

In addition to vision-based tracking, acoustic techniques such as the Acoustic Tag System (ATS) are also used for fish tracking. ATS involves attaching acoustic tags to fish, emitting unique acoustic signals detected by hydrophone receivers. The position of the tagged fish can be estimated using the time difference of arrival of the acoustic signals at multiple receivers, allowing for 3D tracking of fish movement in natural habitats \cite{wu2020integrated}. This section mainly analyzes the relevant literature on fish tracking methods based on visual technology (as shown in Table \ref{tab: tracking}) and acoustic techniques in recent years and provides a systematic summary.

\begin{table*}[]
\begin{threeparttable}
\caption{Summary of different methods in fish tracking.}
\centering
\label{tab: tracking}
\begin{tabular}{@{}ccccccccc@{}}
\toprule

\begin{tabular}[c]{@{}c@{}}Study \\ Site\end{tabular}                & \begin{tabular}[c]{@{}c@{}}Maxmum Fish \\ Amounts\end{tabular} & Points                                                                                              & \begin{tabular}[c]{@{}c@{}}Detection \\ Methods\end{tabular}                  & \begin{tabular}[c]{@{}c@{}}Tracking \\ Methods\end{tabular}                                 & \begin{tabular}[c]{@{}c@{}}Tracking \\ Metrics\end{tabular}  & Advantages                                                                                                               & Limitation                                                                                                                        & References                      \\ \midrule
\begin{tabular}[c]{@{}c@{}}Tank\end{tabular}                & 13                                                      & \begin{tabular}[c]{@{}c@{}}Fish \\ head\end{tabular}                                              & YOLOv2                                                                       & \begin{tabular}[c]{@{}c@{}}Kalman \\ Filter\end{tabular}                                 &                                                \begin{tabular}[c]{@{}c@{}}CIR\tnote{1} \\ CTR\end{tabular}              & \begin{tabular}[c]{@{}c@{}}High frame rate \\ not necessary\end{tabular}                                                 & \begin{tabular}[c]{@{}c@{}}Larger fish \\ quantities \\ increase \\ identification \\ losses\end{tabular}                         & \cite{barreiros2021zebrafish}        \\
% Tank                                                            & 11                                                      & \begin{tabular}[c]{@{}c@{}}Fish \\ center \\ points\end{tabular}                                    & YOLOV3                                                                        &                  \begin{tabular}[c]{@{}c@{}}Euclidean   \\ Distance\end{tabular}                                                                                  & -                                                               & \begin{tabular}[c]{@{}c@{}}Enhance target \\ detection in \\ unclear water\end{tabular}                                  & \begin{tabular}[c]{@{}c@{}}Fish numbers are \\ too small and \\ done in the \\ laboratory\end{tabular}                            & \cite{wageeh2021yolo}                  \\
Coast                                                                &  -                                                        & \begin{tabular}[c]{@{}c@{}}Center of \\ the fish \\ head\end{tabular}                               & \textbf{YOLOv4}                                                                        & \begin{tabular}[c]{@{}c@{}}\textbf{Kalman} \\ \textbf{Filter}\end{tabular}                                 & MOTA\tnote{2}                                               & \begin{tabular}[c]{@{}c@{}}Real-time \\ tracking with \\ high accuracy\end{tabular}                                      & \begin{tabular}[c]{@{}c@{}}Accuracy affected \\ by different sea \\ areas\end{tabular}                                            & \cite{liu2021multi}                  \\
\begin{tabular}[c]{@{}c@{}} Tank\end{tabular}              & 5                                                       & \begin{tabular}[c]{@{}c@{}}Fish head \\ and center \\ of the fish \\ body\end{tabular}              & \begin{tabular}[c]{@{}c@{}}Background \\ subtraction\end{tabular}             & \begin{tabular}[c]{@{}c@{}}Kalman \\ Filter\end{tabular}                                 & \begin{tabular}[c]{@{}c@{}}CIR \\ CTR\end{tabular}              & \begin{tabular}[c]{@{}c@{}}Accurate, fast, \\ and \\ computational \\ inexpensive\end{tabular}                           & \begin{tabular}[c]{@{}c@{}}Fail to predict the \\ motion state of \\ rapidly \\ transitioning\end{tabular}                        & \cite{wang2021parallel} \\
Tank                                                            & 20                                                      & \begin{tabular}[c]{@{}c@{}}Fish head \\ and fish \\ body\end{tabular}                               & \begin{tabular}[c]{@{}c@{}}Background \\ subtraction\end{tabular}             & \begin{tabular}[c]{@{}c@{}}Kalman \\ Filter\end{tabular}                                                                              & \begin{tabular}[c]{@{}c@{}}CIR \\ CTR\end{tabular}              & \begin{tabular}[c]{@{}c@{}}Smoother \\ resulting \\ trajectory\end{tabular}                                              & \begin{tabular}[c]{@{}c@{}}Lower frame \\ speeds lead to \\ more track breaks \\ and higher \\ misidentification\end{tabular}     & \cite{wang2016automated}               \\
Tank                                                            & 5                                                       & Centroid                                                                                            & \begin{tabular}[c]{@{}c@{}}Background \\ subtraction\end{tabular}             & \begin{tabular}[c]{@{}c@{}}Kalman \\ Filter\end{tabular}                                                                              & \begin{tabular}[c]{@{}c@{}}CIR \\ CTR\end{tabular}              & \begin{tabular}[c]{@{}c@{}}Enhances \\ tracking \\ performance \\ under occlusion \\ conditions\end{tabular}             & \begin{tabular}[c]{@{}c@{}}Abnormal water \\ quality leads to an \\ increased chance \\ of fish body \\ overlap\end{tabular}      & \cite{zhao2019algorithm}               \\
% Tank                                                            & -                                                       & Centroids                                                                                           & Otsu                                                                          & \begin{tabular}[c]{@{}c@{}}Manhattan \\ Distance\end{tabular}                               & -                                                               & \begin{tabular}[c]{@{}c@{}}Low cost and \\ removable \\ installation\end{tabular}                                        & \begin{tabular}[c]{@{}c@{}}Stationary fish \\ mistaken for \\ debris or dead\end{tabular}                                         & \cite{soltanzadeh2020prototype}        \\
\begin{tabular}[c]{@{}c@{}}Tank\end{tabular}              & 25                                                      & Fish head                                                                                           & DOH\tnote{3}                                                                & CNN                                                                            & Recall                                                          & \begin{tabular}[c]{@{}c@{}}Corrects \\ trajectory \\ errors, fills \\ gaps, and \\ evaluates \\ credibility\end{tabular} & \begin{tabular}[c]{@{}c@{}}Easy affected by \\ floating objects, \\ ripple reflections, \\ fish sharp turns\end{tabular}          & \cite{xu2017zebrafish}                 \\
Tank                                                            & 10                                                      & \begin{tabular}[c]{@{}c@{}}Head \\ feature \\ point and \\ central \\ feature \\ point\end{tabular} & \begin{tabular}[c]{@{}c@{}}Background \\ subtraction\end{tabular}             & \begin{tabular}[c]{@{}c@{}}Feature point \\ matching\end{tabular}                           & \begin{tabular}[c]{@{}c@{}}Precision \\ Recall\end{tabular}     & \begin{tabular}[c]{@{}c@{}}Two-feature \\ point model \\ reduces \\ tracking \\ difficulty\end{tabular}                  & \begin{tabular}[c]{@{}c@{}}Only traces a few \\ objects for a very \\ short process\end{tabular}                                  & \cite{qian2017feature}                 \\
\begin{tabular}[c]{@{}c@{}}Glass \\ Aquarium\end{tabular}            & 5                                                       & \begin{tabular}[c]{@{}c@{}}The head \\ and tail of \\ fish\end{tabular}                             & \begin{tabular}[c]{@{}c@{}}Adaptive \\ thresholding \\ algorithm\end{tabular} & GNN\tnote{4}                                                                                         & \begin{tabular}[c]{@{}c@{}}Tracking \\ errors\end{tabular}      & \begin{tabular}[c]{@{}c@{}}Accurate \\ tracking by \\ pose constraint, \\ even at high \\ speed\end{tabular}             & \begin{tabular}[c]{@{}c@{}}Unable to handle \\ fish occlusion or \\ attaching\end{tabular}                                        & \cite{xia2016posture}                  \\
\begin{tabular}[c]{@{}c@{}}Fringing \\ Reef, \\ Red Sea\end{tabular} & 4                                                       & \begin{tabular}[c]{@{}c@{}}Fish's \\ body\end{tabular}                                              & \begin{tabular}[c]{@{}c@{}}Fast-RCNN, \\ Inceptiont \\ V2\end{tabular}        & \begin{tabular}[c]{@{}c@{}}Linking \\ consecutive \\ frames\end{tabular}                    & \begin{tabular}[c]{@{}c@{}}3D \\ detection \\ rate\end{tabular} & \begin{tabular}[c]{@{}c@{}}Cost-effective, \\ automated 2D \\ track \\ Reconstruction\end{tabular}                       & \begin{tabular}[c]{@{}c@{}}Small groups of \\ fish studied\end{tabular}                                                           & \cite{engel2021situ}                   \\
Tank                                                                 & $50$                                                  & Head                                                                                                & ResNet-101                                                                    & \begin{tabular}[c]{@{}c@{}}Mahalanobis \\ distance and \\ cosine \\ similarity\end{tabular} & \begin{tabular}[c]{@{}c@{}}MOTA \\ IDF1\tnote{5}\end{tabular}            & \begin{tabular}[c]{@{}c@{}}Performance \\ well under \\ multiple \\ negative factors\end{tabular}                        & \begin{tabular}[c]{@{}c@{}}Bad performance \\ of long-term \\ tracking\end{tabular}                                               & \cite{li2022cmftnet} \\
\begin{tabular}[c]{@{}c@{}}Tank \\ Pond\end{tabular}                 & $50$                                                  & Body                                                                                                & \textbf{Transformer}                                                                   & \begin{tabular}[c]{@{}c@{}}\textbf{Hungarian} \\ \textbf{algorithm}\end{tabular}                              & \begin{tabular}[c]{@{}c@{}}MOTA \\ IDF1\end{tabular}            & \begin{tabular}[c]{@{}c@{}}Accommodates \\ individuals with \\ significant \\ appearance \\ variations.\end{tabular}     & \begin{tabular}[c]{@{}c@{}}Limitations in \\ accurate ID \\ matching at high \\ stocking densities \\ (over 50 fish)\end{tabular} & \cite{li2024tfmft}                     \\
Tank                                                                 & 8                                                       & Head                                                                                                & LSTM                                                                          & \begin{tabular}[c]{@{}c@{}}Kalman \\ Filter\end{tabular}                                 & \begin{tabular}[c]{@{}c@{}}Precision \\ Recall\end{tabular}     & \begin{tabular}[c]{@{}c@{}}Cross-view \\ more robust in \\ high densities\end{tabular}                                   & \begin{tabular}[c]{@{}c@{}}Multi -view map \\ matching is \\ difficult, and the \\ calculation \\ amount is large\end{tabular}    & \cite{wang20173d}                      \\
Tank                                                                 & 49                                                      & Head                                                                                                & \begin{tabular}[c]{@{}c@{}} DoH \\ CNN\end{tabular}               & \begin{tabular}[c]{@{}c@{}}Iterative \\ tracking \\ strategy\end{tabular}                   & \begin{tabular}[c]{@{}c@{}}Precision \\ Recall\end{tabular}     & \begin{tabular}[c]{@{}c@{}}Tracking \\ individuals \\ exhibiting \\ frequent \\ occlusions\end{tabular}                  & \begin{tabular}[c]{@{}c@{}}Requires \\ individuals to \\ have at least one \\ body part that \\ remains robust\end{tabular}       & \cite{wang2017robust}                  \\ \bottomrule
\end{tabular}
        \begin{tablenotes}
            \item[1]  CTR (correct tracking rate), CIR (correct identification reason)
            \item[2] MOTA (multiple objects tracking accuracy) (as shown in formula \ref{fumula:Mota})
            \item[3] DOH (determinant of Hessian)
            \item[4] GNN (global nearest neighbour)
            \item[5] IDF1(identification-score, as shown in formula \ref{fumula:idf1})
        \end{tablenotes}
\end{threeparttable}
\end{table*}

\subsection{Fish tracking based on 2-dimensional visual information}
Fish tracking methods can be broadly categorized into three main approaches: classical algorithms, kernel correlation filter-based algorithms, and deep learning-based tracking algorithms \cite{mei2022recent}. Each category encompasses various techniques with their strengths and limitations, which will be explored in more detail in the following subsections.

\subsubsection{Fish tracking based on classical algorithms}

Classical algorithms have been widely used to address the challenges of fish tracking in complex underwater environments, such as rapid posture changes, occlusion, overlap, and poor image quality. The Tracking-Learning-Detection (TLD) algorithm, which updates salient features and target model parameters through online learning, has shown promise in providing stable tracking \cite{kalal2011tracking}. However, its median-flow tracker may fail when fish change their swimming posture rapidly. However, an adaptive scale mean-shift (ASMS) algorithm, utilizing fish shape and colour features, can handle posture changes, uneven illumination, and complex backgrounds \cite{wang2019fish}. 

Preserving individual fish identities during occlusion and overlap remains a significant challenge. Techniques that extract head shape or body geometry features have been explored \cite{terayama2015appearance, terayama2016multiple}, but their effectiveness may be limited by the rapid movement and intense geometry of fish bodies \cite{soltanzadeh2020prototype}. Adaptive thresholding algorithms, which estimate each pixel's threshold based on its adjacent region, have shown promise in segmenting individual fish in binarized images \cite{qian2016effective}. The global nearest neighbour algorithm with fish posture as a tracking constraint has been used to track small numbers of zebrafish \cite{xia2016posture}, but it lacks individual recognition ability, leading to track exchanges during overlap or occlusion. The Toxld algorithm addresses this issue using intensity histograms and Hu-moments to link trajectory fragments and preserve individual fish identities \cite{rodriguez2017toxid}. However, with this method, the error increases with the number of fish.

To deal with poor image quality, retinex (MSR) based enhancement algorithms combined with object detection have been used to improve fish detection in unclear underwater images \cite{wageeh2021yolo, anas2020detecting, mohamed2020msr}. Kalman filters may not always be optimal for underwater fish tracking due to the presence of non-Gaussian noise and complex environments \cite{qian2014automatically}. Mean offset technology, which models fish probability density based on colour histograms, can fail when the background colour closely resembles the fish colour distribution. Tracking algorithms based on covariance matrices of pixel-based feature sets, incorporate spatial and statistical characteristics, making them more suitable for tracking fish in challenging underwater environments \cite{spampinato2014rule, spampinato2014understanding}.

\subsubsection{Fish tracking based on Kalman filters}

\textcolor{black}{The Kalman filter is an efficient autoregressive filter, which can estimate the state of a dynamic system from sensory measurements taken from an environment with uncertainties posed by noise and interferences \cite{chen2011kalman}. However, when applied alone to complex tracking scenarios, it faces challenges in handling occlusions and maintaining accurate tracking in high-density situations \cite{wang2016automated}. 
% Wang et al. \cite{wang2016automated} demonstrated its effectiveness for tracking fish heads and fish bodies, e.g. by employing rectangle chains. While their results showed successful occlusion detection in low fish density scenarios when fish heads remained visible, the approach struggled with the correction of incorrect detections in high-density situations.
To address these limitations, researchers explored combining Kalman filtering with advanced object detection methods. You Only Look Once (YOLO) \cite{mathias2022occlusion} has emerged as a leading detection method due to its real-time performance and accuracy in identifying fish within video frames. Building upon these foundations, the Simple Online and Realtime Tracking (SORT) algorithm \cite{zhao2019algorithm} integrates YOLO-based detection results with Kalman filtering for motion prediction and employs the Hungarian algorithm for data association \cite{bewley2016simple}. This integration enables high-speed tracking (260 FPS) by focusing solely on bounding box information, making it ideal for applications requiring fast and efficient tracking. However, the performance of SORT is limited by its reliance on pure detection results without considering object appearance features, making it vulnerable to tracking failures during occlusions or complex scenarios \cite{pereira2022sort}.}
% Building upon the Kalman filter, the Simple Online and Realtime Tracking (SORT) algorithm emerged as a simple yet effective multi-target tracking approach \cite{zhao2019algorithm}. SORT is a real-time object-tracking algorithm that combines Kalman filtering for motion prediction with a Hungarian algorithm for data association \cite{bewley2016simple}. It achieves high-speed tracking (260 FPS) by focusing only on bounding box information without using appearance features, making it ideal for applications requiring fast and efficient tracking.}
% SORT utilizes a Kalman filter for frame-by-frame data correlation and the Hungarian algorithm for correlation measurement \cite{bewley2016simple}.  
% While this algorithm performs well at high frame rates, it has limitations in exploiting features from object surface, making the tracking performance heavily dependent on detection results \cite{pereira2022sort}.

\begin{figure}[t]
\centering
\includegraphics[width=9cm]{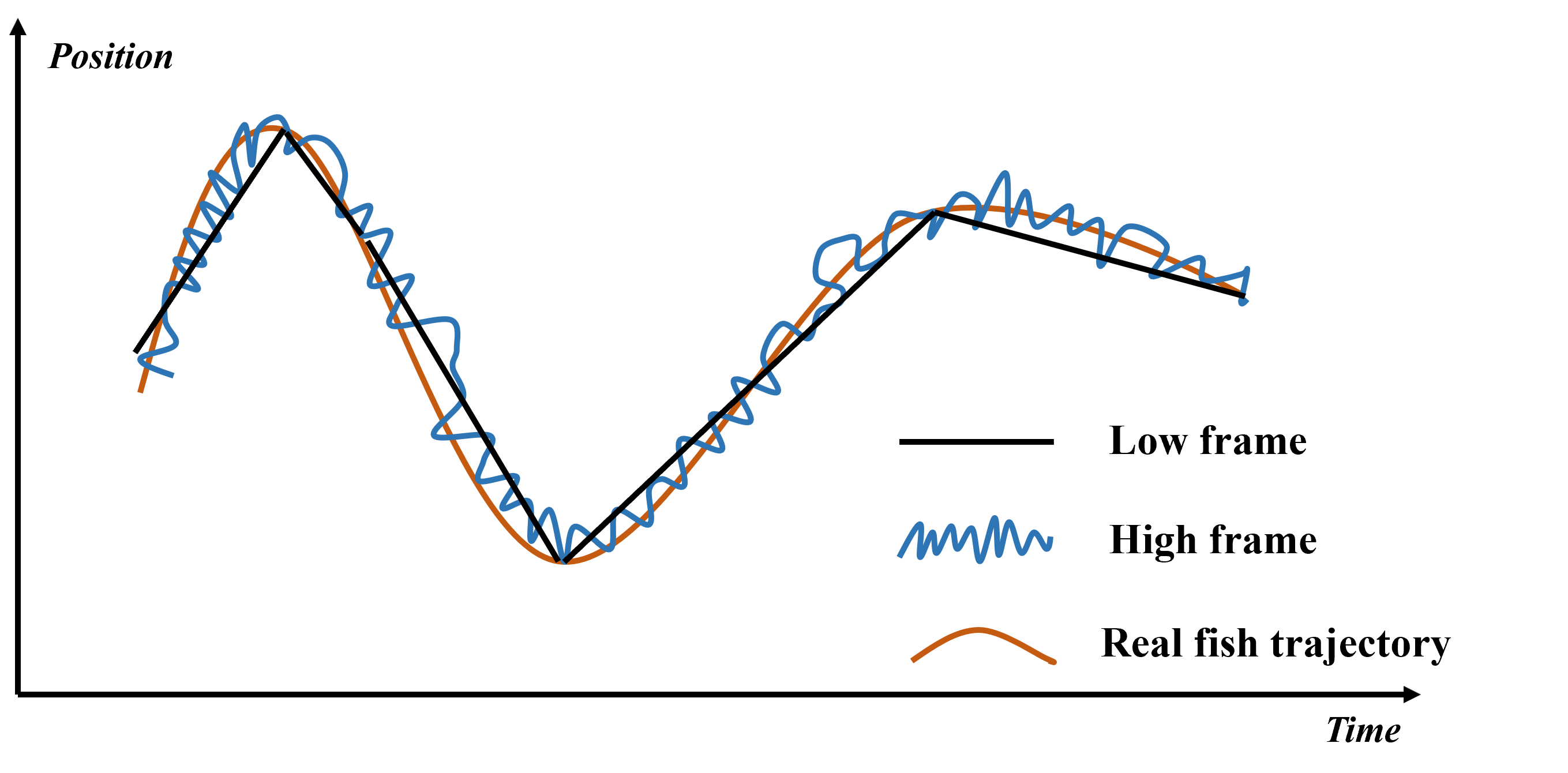}
\caption{Fish trajectory under different frame rates (This figure was adapted from \cite{delcourt2013video}).}
\label{fig: frames}
\end{figure}

\textcolor{black}{To address these limitations, DeepSORT was developed by incorporating a convolutional neural network (CNN) model trained on large-scale datasets \cite{wojke2017simple}. This advancement significantly improved tracking robustness to target loss and occlusions, making it particularly suitable for aquaculture applications \cite{barreiros2021zebrafish}. For instance, Mathias et al. \cite{mathias2022occlusion} successfully demonstrated this capability by implementing a hybrid adaptive DeepSORT with YOLOv3 \cite{farhadi2018yolov3} for underwater object tracking.
% To address these limitations, DeepSORT was developed, incorporating a convolutional neural network (CNN) model trained on large-scale datasets to enhance feature extraction and improve its robustness to target loss and occlusions \cite{wojke2017simple}. Recent literature indicates that YOLO and DeepSORT algorithms are gaining popularity for fish tracking in aquaculture applications \cite{barreiros2021zebrafish}. For example, Mathias et al. \cite{mathias2022occlusion} used a hybrid adaptive deep SORT and YOLOv3 approach for tracking objects in underwater scenarios, with the presence of occlusions. 
However, challenges arise when fish undergo rapid body shape changes during fast turns, leading to blurry and difficult-to-track images \cite{wang2021parallel}. To mitigate this issue, shorter exposure times and boundary boxes of variable size can be used, with the boundary boxes being estimated according to the motion state. Additionally, frame rate optimization has proven crucial for Kalman filter-based tracking performance: higher rates facilitate more linear fish motion predictions, while lower rates increase misidentification risks (as shown in Fig. \ref{fig: frames}).}

\textcolor{black}{Recent advancements in tracking algorithms have introduced several innovative approaches to address challenges in occlusions, rapid movements, and varying environmental conditions. One major category focuses on feature enhancement strategies: StrongSORT improved upon DeepSORT by incorporating stronger appearance features through a more sophisticated feature extractor based on ResNet-50 backbone networks and enhanced motion cues with Kalman filter refinements \cite{du2023strongsort}. Its integration with GN-YOLOv5 \cite{tang2024hic} further optimized both performance and processing speed through better feature extraction and more accurate object detection \cite{zhai2023multi}.}
\textcolor{black}{Another category advances association strategies: BoT-SORT innovated tracking through a redesigned association mechanism that combines cosine distance metrics (comparing new detections' features with stored tracklet features) for appearance-matching and Camera Motion Compensation (CMC) for handling camera movements \cite{aharon2022bot}. Unlike StrongSORT which primarily relies on appearance features, BoT-SORT introduced a hybrid association strategy that balances appearance similarity with motion consistency. This dual approach has proven particularly effective in challenging scenarios like aquaculture, where combined with YOLOv8 \cite{sohan2024review}, it successfully tracks fast-moving fish despite rapid movements and appearance changes by better leveraging both motion and appearance information \cite{xing2024sonar}.}

\textcolor{black}{The third category introduces observation-centric (OC) approaches, where OC-SORT and Deep OC-SORT target dense population scenarios by prioritizing actual detection results over predictions, unlike traditional trackers that rely heavily on predictive models \cite{cao2023observation, maggiolino2023deep}.
OC-SORT prioritizes high-quality observations and introduces an observation-centric Kalman filter (OCKF) for more reliable state estimation, along with a new track recovery strategy and adaptive track confidence mechanism \cite{cao2023observation}. Deep OC-SORT further enhances this approach by incorporating a deep association module for better feature matching, an enhanced track rebirth mechanism, and more sophisticated appearance modelling \cite{maggiolino2023deep}. Building on these approaches, ByteTrack pioneered a detection utilization strategy that processes both high and low-confidence detections through separate association pipelines \cite{zhang2022bytetrack}. This differs from previous approaches that typically discarded low-confidence detections, enabling ByteTrack to maintain tracking continuity even under challenging conditions. Its success in aquaculture applications \cite{alaba2024multifish, zhao2024fish} demonstrates robust performance under varying lighting conditions while effectively reducing object loss and trajectory fragmentation, with its flexible architecture supporting various model sizes for different application scenarios \cite{qian2023fish}.}

\textcolor{black}{The evolution of tracking algorithms from SORT to recent approaches like StrongSORT, BoT-SORT, OC-SORT, and ByteTrack offers diverse solutions for different tracking scenarios. For practical applications, StrongSORT is suitable when strong feature extraction is crucial, BoT-SORT works well with moving cameras and appearance changes, while OC-SORT and ByteTrack are particularly effective in crowded scenes. Future research could focus on combining these strengths, such as integrating robust feature extraction with observation-centric approaches, while also addressing computational efficiency to enable real-time tracking in resource-constrained environments.}

\subsubsection{Fish tracking based on deep learning}
% \textcolor{black}{The deep learning based models have also been applied to fish tracking. At present, the main method of fish tracking is based on TBD (tracking by detection) \cite{kalal2011tracking}. However, occlusion between fishes is the biggest challenge of TBD tracking methods. To solve the problem caused by occlusions, existing methods assign the tracker with the detected objects to generate numerous trajectory fragments, and then post-process these trajectory fragments and link them together \cite{bhateja2020suze}. }
\textcolor{black}{Deep learning has revolutionized fish tracking approaches, introducing various methodologies to address the complex challenges in aquaculture environments. These approaches can be broadly categorized into three main streams: (1) identity-based methods like idTracker \cite{perez2014idtracker} that focus on extracting unique fingerprint features for individual fish recognition, (2) detection-tracking frameworks that combine CNN-based detection with sophisticated tracking mechanisms, and (3) advanced tracking architectures, such as Siamese networks that focus on similarity learning between frames and transformers that leverage self-attention mechanisms. While these methods have shown significant improvements over traditional approaches, they each face distinct challenges in handling occlusions, maintaining tracking stability in dense populations, and adapting to varying environmental conditions \cite{kalal2011tracking, bhateja2020suze}.}

Notable identity-based tracking algorithms such as idTracker \cite{perez2014idtracker} and its upgraded version, idtracker.ai \cite{romero2019idtracker}, have demonstrated success in individual fish tracking. idTracker pioneers the concept of extracting unique visual fingerprints for each animal, utilizing image-based features to maintain individual identities throughout videos. The upgraded version, idTracker.ai, enhances this approach by incorporating deep learning techniques. These algorithms extract unique fingerprint features from each animal and identify targets throughout the video, enabling automated tracking of untagged animals within groups. However, their application is often limited to controlled environments where fish movement is restricted to prevent overlapping, making them less suitable for real-world 3D tracking scenarios.
% To address the occlusion issue, several notable multi-target tracking algorithms have been proposed, such as idTracker \cite{perez2014idtracker} and its upgraded version, idtracker.ai \cite{romero2019idtracker}. These algorithms extract unique fingerprint features from each animal in a set of videos and then identify each target in the video, enabling the tracking of individuals within a group by automatically identifying untagged animals. Although these methods have been widely used for tracking juvenile fish and small animals, the experimental setup restricted fish from swimming up and down to avoid overlapping, simplifying the task compared to real-world 3D tracking scenarios.

Detection-based tracking frameworks have achieved significant advancements by combining CNN-based detection methods with complementary tracking techniques. These approaches integrate with other techniques, such as head detection, motion state prediction, and SVM-based verification \cite{wang2017robust}, showing improved robustness compared to idTracker in scenarios with higher fish density and increased occlusion frequency \cite{xu2017zebrafish}. 
% Further advancements in deep learning-based fish tracking have been achieved by combining CNN-based methods with other techniques, such as head detection, motion state prediction, and verification using support vector machine (SVM) based classifiers \cite{wang2017robust}. These approaches have demonstrated more robust tracking performance compared to idTracker when the fish density is higher, and the occlusion frequency increases, highlighting the potential of deep learning in handling complex tracking scenarios \cite{xu2017zebrafish}.
Despite the progress made in controlled laboratory environments, real-world marine environments pose additional challenges for fish tracking, such as light fluctuations and waves. To address these issues, researchers have developed methods like the real-time multi-class fish (RMCF) stock statistics method, which uses YOLOv4 \cite{gai2023detection} as the backbone network and adopts a parallel two-branch structure based on deep learning for detecting fish species, tracking, and counting fish \cite{liu2021multi}. Although these methods have shown promising results in complex marine environments, their recognition accuracy may vary in different sea areas due to differences in colour cast and contrast, necessitating the retraining of the network weight coefficients.

\begin{figure}[t]
\centering
\includegraphics[width=8.5cm]{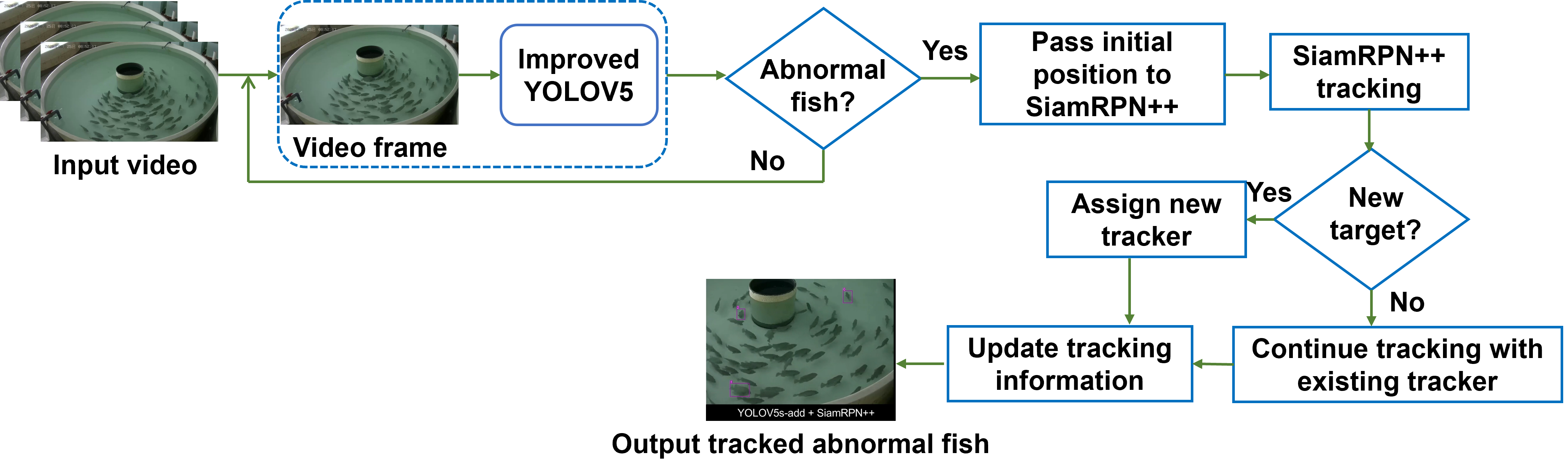}
\caption{Fish tracking method based on YoloV5 and SiamRPN++ (This figure was reproduced from \cite{wang2022real}).}
\label{fig:SamRPN}
\end{figure}

Advanced tracking architectures in fish tracking have evolved along two main directions: Siamese networks and transformer-based methods \cite{chicco2021siamese, vaswani2017attention}. Siamese network trackers have gained attention in recent years due to their exceptional tracking speed and high accuracy. The introduction of advanced algorithms, such as SiamRPN++ (as shown in Fig. \ref{fig:SamRPN}), has further demonstrated the performance of Siamese networks, surpassing the performance of the tracking algorithms based on correlation filters \cite{li2019siamrpn++}, \cite{wang2022real}. Although there are currently few articles on Siamese networks specifically for fish tracking, this approach has the potential to be used in the field.

The emergence of transformer-based tracking methods has revolutionized object tracking, including fish tracking in aquaculture environments. Initially proposed for natural language processing tasks, transformers have been successfully adapted for computer vision tasks, including object detection and tracking \cite{carion2020end}. Transformer-based trackers, such as TransTrack \cite{chen2021transformer} and STARK \cite{yan2021learning}, have demonstrated state-of-the-art performance on various tracking benchmarks. \textcolor{black}{In the context of fish tracking, transformer-based methods have shown promising results. For instance, Li et al. \cite{li2024tfmft} proposed a transformer-based multiple fish tracking model (TFMFT) to address the issue of instance loss in aquaculture ponds with complex background disturbances. This approach represents a shift from previous tracking-by-detection (TBD) methods \cite{kalal2011tracking} that relied on separate detection and embedding paradigms. Instead, the end-to-end Transformer architecture enables unified processing for both training and tracking. Building upon this work, Liu et al. \cite{liu2024fishtrack} introduced FishTrack, an online multi-fish tracking model, that incorporates a specially designed encoder to encode historical position information of fish and automatically update spatiotemporal information in an auto-regressive manner. This approach allows for the fusion of spatiotemporal information of fish targets without manual feature selection, demonstrating robustness even under long-term occlusion conditions.}

\textcolor{black}{Despite these advancements, significant challenges remain across different tracking approaches. Identity-based methods like idTracker.ai \cite{romero2019idtracker} still face limitations in handling complex 3D environments and high-density scenarios. Detection-tracking frameworks struggle with environmental variations and often require retraining for different conditions, while Siamese networks, although promising, need further validation in real-world aquaculture settings. Transformer-based methods face computational complexity issues and require substantial training data, which can be scarce in aquaculture settings. They may also struggle with extended tracking sequences and lack inherent temporal modelling capabilities. Specific challenges include achieving accurate ID matching at high stocking densities and maintaining tracking stability during long-term monitoring. Furthermore, the potential for overfitting when the training data is limited or not diverse enough is a particular concern in aquaculture, where environmental conditions can vary widely. For example, TFMFT faces limitations in achieving accurate identity (ID) matching at high stocking densities and experiences some tracking losses during long-term monitoring.}

\textcolor{black}{Future research should focus on further adapting transformer architectures to address specific challenges, such as developing efficient training strategies to handle the limited availability of annotated fish-tracking datasets, improving performance in high-density scenarios, and enhancing long-term tracking stability. Additionally, efforts should be directed towards developing more efficient and lightweight models, improving their ability to handle long-term tracking and occlusions, and creating strategies to effectively train these models with limited data.}

% \textcolor{black}{Despite these advancements, challenges remain. TFMFT, for example, faces limitations in achieving accurate ID matching at high stocking densities and experiences some tracking losses during long-term monitoring. As transformer-based methods continue to evolve, they are expected to play an increasingly important role in fish-tracking applications. Future research should focus on further adapting transformer architectures to address the specific challenges of underwater environments in aquaculture. This includes developing efficient training strategies to handle the limited availability of annotated fish-tracking datasets, improving performance in high-density scenarios, and enhancing long-term tracking stability.}

\subsection{Fish tracking based on 3-dimensional visual information}

3-D tracking methods offer advantages over 2-D tracking algorithms, as they can be used to study the behaviour of social animals and effectively address occlusion problems. However, 3-D tracking also presents significant challenges due to the large number of fish, similar individual appearance, occlusion, and uncertainty of stereo matching.

\begin{figure}[t]
\centering
\includegraphics[width=8.5cm]{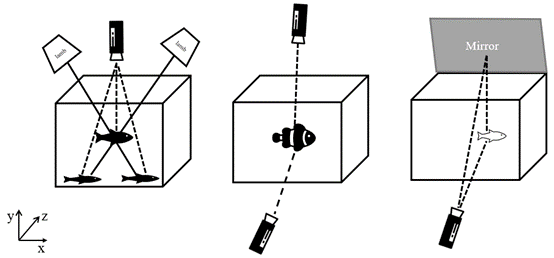}
\caption{Three methods to measure the 3-D position of a fish in an aquarium (This figure
was adapted from \cite{delcourt2013video}).}
\label{fig:3dtracking}
\end{figure}

Two main types of 3-D tracking methods have been developed: ``shadow" and ``stereo" methods (as shown in Fig. \ref{fig:3dtracking}). The ``shadow" method, which requires only one camera, uses the shadow of the fish projected onto the substrate as a second view of the shoal. By calculating the 2-D positions of the fish and its shadow, the 3-D position of the fish can be obtained through triangulation. However, this method becomes increasingly difficult as the number of fish increases and shadows may be obscured, as it requires detecting each fish and its corresponding shadow \cite{palconit2021three}.

Stereoscopic methods use multiple cameras, a camera and a mirror to capture simultaneous images at different angles \cite{delcourt2013video}. Some researchers have developed platforms that use a single camera and mirror to obtain 3-D coordinates of fish \cite{mao2016research, xiao2016research}. These methods calculate the centre coordinates of fish and combine the association of mirror view and direct view for tracking, addressing the problem of target loss caused by occlusion. However, they require high-precision equipment and may suffer from correspondence deviations due to the pixel centres of real and virtual fish not being at the same point. Moreover, these methods are not suitable for actual production environments.

\begin{figure}[t]
\centering
\includegraphics[width=8.5cm]{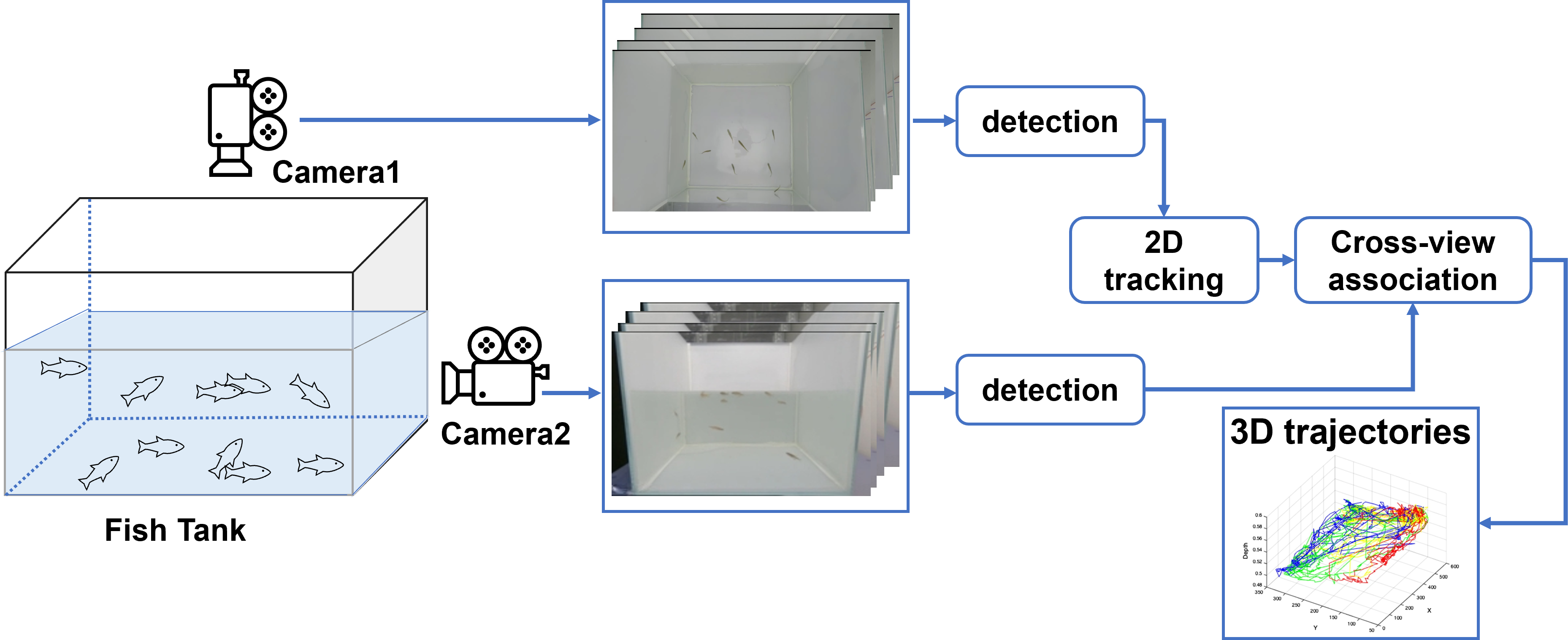}
\caption{Three-dimensional trajectory of multiple fish in water tank via multi video tracking (This figure was inspired by \cite{cheng2018obtaining} with minor changes).}
\label{fig: 3dTrajectory}
\end{figure}

In theory, two cameras are sufficient for stereo imaging. 3-D tracking with two cameras involves obtaining the 2-D motion trajectory from the top view with a larger viewing angle and then performing 3-D matching of the top view tracking results with the feature points in the side view to obtain movement of the object in 3-D space (as shown in Fig. \ref{fig: 3dTrajectory}) \cite{cheng2018obtaining}. \textcolor{black}{To track many objects, three or more high-speed cameras usually are needed to capture synchronous videos to mitigate the ambiguities and errors in distinguishing the objects. For example, Wang et al. \cite{wang20163d} determined the location of fisheye under the top and side views using mixed Gaussian and Gabor models, respectively, and then obtained their 3-D motion trajectories by associating the top-view tracking results with the trajectories of two side views \cite{wang20173d}. However, this method is limited due to the difficulty in distinguishing the characteristics from the eye area of fish. Furthermore, analyzing fish movement behaviour in three views requires the installation of complex equipment and the association and stereo matching between the views \cite{liu2019video}.}

Occlusion remains one of the main challenges in 3D fish tracking, as it is in other Multiple Object Tracking (MOT) tasks. However, the frequency of occlusion has not been adequately measured in the current literature, with the complexity indicator of the datasets used in existing studies typically being the number of fish rather than an assessment of fish occlusion events. For instance, a demo video in \cite{qian2017feature} shows only 4 occlusion events within 15 seconds for a group of 10 fish. Current system evaluations assess parameters such as ID swaps, fragments, precision, and recall for the generated 2-D and 3D tracks without describing how these indicators are calculated. The lack of uniform indicators makes it difficult to fairly compare the methods presented in various studies. Furthermore, most of the literature does not provide open-source code and annotated data, limiting the reproducibility of the results.
A recent study by \cite{pedersen20203d, li2022cmftnet} introduced a standard MOT evaluation framework for fish tracking, providing a good model for multi-target fish tracking. A unified evaluation standard should be introduced to ensure the fairness of fish tracking comparisons and facilitate progress in this field.

\begin{figure}[h]
    \centering
    \includegraphics[width=\linewidth]{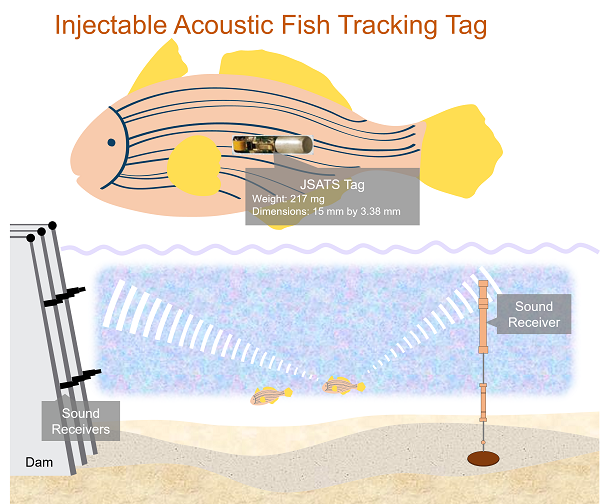}
    \caption{Fish tracking method based on an acoustic tag (This figure was reproduced from\cite{martinez2021large} with minor changes).}
    \label{fig: acceleration sensor}
\end{figure}

\subsection{Fish tracking based on acoustic tag system }

The Acoustic Tag System (ATS), a passive acoustic monitoring technology, has become an important means of monitoring fish trajectories and studying fish behaviour \cite{duzhuang2024recent}. Unlike vision-based tracking methods, which rely on clear water conditions and sufficient lighting, ATS can provide reliable tracking data in challenging underwater environments, such as turbid waters or low-light conditions \cite{pursche2014evaluation}.
The appropriate type and parameters of the acoustic tag (also called an acoustic signal transmitter) are selected according to the size of the fish and the research period (as shown in Fig. \ref{fig: acceleration sensor}) \cite{matley2022global}. \textcolor{black}{The applications of ATS are diverse, including fish resource abundance assessment, swimming pattern analysis, habitat evaluation, spawning site identification, survival rate estimation, and behavioural differentiation \cite{aspillaga2021high, lennox2017envisioning}. Kolarevic et al. \cite{kolarevic2016use} demonstrated the use of acoustic acceleration transmitter tags for monitoring Atlantic salmon swimming activity in recirculating aquaculture systems (RAS), while Fore et al. \cite{fore2017biomonitoring} found it feasible to monitor fish depth in real-time in commercial fish farms using acoustic telemetry.}

\textcolor{black}{Acoustic tag systems have been used to monitor fish movement and trajectories, obtain the three-dimensional coordinates of the fish in real-time, and perform related data analysis and application. A notable example of an ATS application involves 682 subyearling Chinook salmon tagged with injectable acoustic transmitters in the Snake River, Washington State \cite{martinez2021large}. This study generated over 5 million clean detections and 403,900 3-D trajectories, helping to assign fish to passage routes through a dam and expanding understanding of near-dam fish behaviour \cite{martinez2021large}. Leclercq et al. \cite{leclercq2018application} and Munoz et al. \cite{munoz2020acoustic} evaluated the feasibility of using passive acoustic telemetry to monitor the welfare of fish in sea cages on an industrial scale, and the experimental results showed that the acoustic tracking system proved to be an effective tool for fish behaviour analysis. However, the application of acoustic tags in commercial aquaculture faces significant challenges. Macaulay et al. \cite{macaulay2021tag} systematically reviewed studies using tags to monitor farmed fish behaviour and found concerning results regarding tag-associated mortality. Mortality was 10 times higher in marine caged fish compared to tanks. Moreover, mortality in marine caged fish increased significantly in longer trials, ranging from 4\% in single-day trials to 36\% after 100 days.}

\textcolor{black}{Despite these challenges, ATS offers unique advantages in monitoring fish movement and obtaining real-time three-dimensional trajectory coordinates \cite{macaulay2022passive}. It excels in in-situ observation and simplifies data processing over vision-based methods. However, the technology's reliance on tags attached to fish can potentially impact on fish health and survival \cite{klinard2020living, notte2022application}, necessitating careful monitoring and prompt data analysis. Future research should focus on improving tag design, developing standardized reporting protocols, investigating mortality causes in marine caged fish, and exploring less invasive alternatives. By addressing these challenges, ATS could become a more viable tool for understanding and improving fish welfare and productivity in aquaculture, though its use should be limited currently and results interpreted cautiously in commercial settings.}

\subsection{Tracking evaluation metrics}
Multi-target tracking evaluation indices directly reflect an algorithm's tracking ability. The MOT Challenge official multi-objective tracking evaluation indicators \cite{dendorfer2021motchallenge} provide a standardized framework for performance assessment. Key metrics include Multiple Object Tracking Accuracy (MOTA) and Multiple Object Tracking Precision (MOTP).

The MOTA combines three sources of errors to evaluate a tracker’s performance, as follows:
\begin{equation}
\label{fumula:Mota}
M O T A=1-\frac{\Sigma_t\left(F N_t+F P_t+I D_{S W_t}\right)}{\Sigma_t G T_t}
\end{equation}
where $F N_t, F P_t, I D_{s W_t}$, and $G T_t$ represent the number of false negatives, false positives, identity switches, and ground truth targets in frame $t$, respectively.

The MOTP is used to measure misalignment between annotated and predicted object locations, defined as:

\begin{equation}
MOTP=\frac{\sum_{i, t} d_t^i}{\sum_t c_t} .
\end{equation}
 where $d_t$ is the distance between the localization of objects in the ground truth and the detection output
$c_t$ is the total matches made between ground truth and the detection output.

Identification-Score ($IDF_1$) comprehensively considers Identification Precision ($IDP$) and Identification Recall ($IDR$) rate:
\begin{equation}
\label{fumula:idf1}
I D F_1=\frac{T P}{T P+0.5 F P+0.5 F N}
\end{equation}
where True Positive ($TP$), False Positive $(F P)$, and False Negative $(F N)$ involved in $IDF_1$ all consider ID, so the indicator is more sensitive to the accuracy of ID information.

To better capture the specific challenges of tracking fish populations, some literature has introduced additional metrics, such as Correct Tracking Ratio (CTR) and Correct Identification Ratio (CIR). CTR measures the percentage of correctly tracked frames for individual fish as follows,
\begin{equation}
C T R=\frac{\sum(\text { NumberOfCorrectFramesOfSingleFish })}{\text { NumberOfFish } \times \text { NumberOfFrames }}
\end{equation}
while $CIR$ represents the probability of correctly identifying all fish after an occlusion event:
\begin{equation}
C I R=\frac{\text { TimesThatAllFishGetCorrectIdentityAfterOcclusion }}{\text { NumberOfOcclusionsEvents }}
\end{equation}

In addition to those metrics, tracking speed is another important factor to consider when evaluating fish-tracking algorithms, especially for real-time applications. Some common metrics for measuring tracking speed include frames per second (FPS) and processing time per frame. FPS indicates the number of frames a tracking algorithm can process in one second while processing time per frame measures the average time taken to process a single frame. Higher FPS and lower processing time per frame are desirable for efficient and real-time tracking performance.
These metrics offer a valuable foundation for evaluating fish tracking performance, comprehensively assessing various errors, fish-specific challenges, and tracking speed. However, they may not always capture the full complexity of fish-tracking scenarios and can be limited by the lack of widespread adoption and the need for detailed ground-truth annotations. There is a potential scope to develop more specialized metrics and evaluation protocols by considering the specific requirements and challenges of fish tracking applications.

\begin{table*}[b]
\caption{Comparison of fish tracking methods}
\label{tab:tranckingcompared}
\begin{tabular}{lllll}
\hline
Method    & Advantages                                                                                                                     & Disadvantages                                                                                                                                               & Welfare Implications                                                                                 & Suitable Environments                                                                                       \\ \hline
\begin{tabular}[c]{@{}l@{}}2D Visual\\ \cite{li2022cmftnet,mandel2023detection}\\ \cite{wang2022real,mohamed2020msr}\end{tabular}& \begin{tabular}[c]{@{}l@{}}Non-invasive\\ Cost-effective\\ Real-time tracking\end{tabular}                                & \begin{tabular}[c]{@{}l@{}}Limited to surface or shallow waters\\ Affected by water turbidity\\ Occlusion issues in dense populations\end{tabular} & \begin{tabular}[c]{@{}l@{}}Minimal stress\\ No physical interaction\end{tabular}                     & \begin{tabular}[c]{@{}l@{}}Clear water ponds\\ Shallow tanks\\ Surface monitoring in sea cages\end{tabular} \\ \hline
\begin{tabular}[c]{@{}l@{}}3D Visual\\ \cite{pedersen20203d, wu2024online}\\ \cite{liu2024fishmot,fan2024exploring}\end{tabular} & \begin{tabular}[c]{@{}l@{}}Accurate spatial tracking\\ Works in deeper waters\\ Better for dense populations\end{tabular} & \begin{tabular}[c]{@{}l@{}}Higher computational requirements\\ More complex setup\\ Costly equipment\end{tabular}                                  & \begin{tabular}[c]{@{}l@{}}Minimal stress\\ Potential light disturbance\end{tabular}                 & \begin{tabular}[c]{@{}l@{}}Deep tanks\\ Sea cages\\ Research facilities\end{tabular}                        \\ \hline
\begin{tabular}[c]{@{}l@{}}Acoustic Tags\\ \cite{koeberle2023whole, matley2024trackdat}\\ \cite{lennox2023positioning, kanigan2024acoustic}\end{tabular} & \begin{tabular}[c]{@{}l@{}}Works in turbid waters\\ Long-range tracking\\ Effective in open waters\end{tabular}           & \begin{tabular}[c]{@{}l@{}}Invasive (if using tags)\\ Potential signal interference\\ Higher initial cost\end{tabular}                             & \begin{tabular}[c]{@{}l@{}}Stress from tagging procedure\\ Potential behavioral changes\end{tabular} & \begin{tabular}[c]{@{}l@{}}Open sea farms\\ Large lakes\\ Deep water environments\end{tabular}              \\ \hline
\end{tabular}
\end{table*}

\subsection{Comparative analysis of tracking methods}

\textcolor{black}{2D visual methods are ideal for controlled environments with clear water, offering a balance between cost-effectiveness and non-invasiveness. However, they struggle in turbid or deep waters and occlusions in dense populations. 3D visual tracking provides more comprehensive data but requires more sophisticated equipment and setup, making it more suitable for research-intensive or high-value production environments. Acoustic tag methods excel in challenging water conditions but involve more invasive procedures that may impact fish welfare (as shown in Table \ref{tab:tranckingcompared}). The choice of a method often depends on the specific aquaculture environment. For example, shallow freshwater ponds might benefit more from 2D visual systems, while deep-sea cages may require acoustic or advanced 3D visual setups.}

\textcolor{black}{Regional preferences for tracking methods vary significantly. In Asia, 2D and 3D visual tracking is used predominantly, especially in countries like China and Japan \cite{luo2022rapid, kuroda2023history}. This preference is driven by the prevalence of intensive ponds, tank aquaculture, and strong technological infrastructure. A balanced approach is used in Europe, with significant adoption of all three methods. In North America, the integration of advanced AI with tracking systems is often employed, particularly by combining visual and acoustic tag methods for comprehensive monitoring \cite{matley2022global}. In Oceania, acoustic tracking is favoured due to the prevalence of open-water aquaculture, especially in tuna farming \cite{looby2022quantitative}. Environmental conditions, regulatory frameworks, economic factors, and available technological infrastructure shape the trends for adoption. Developing regions in Africa and parts of Asia often lack access to advanced tracking technologies, highlighting the need for cost-effective solutions for small-scale farmers.}

\textcolor{black}{Animal welfare is an increasingly important consideration for the selection and development of tracking methods \cite{zhang2024advancements}. Visual methods generally cause minimal direct stress but may alter behaviour due to changes in lighting or the presence of equipment \cite{cai2023semi}. Acoustic tagging can cause initial stress during the tagging procedure and may affect long-term behaviour due to the presence of the tag \cite{matley2024trackdat, kanigan2024acoustic}. All methods can potentially cause stress if they require changes to the fish's environment or handling of the fish.
Research on less invasive techniques includes the development of smaller, lighter acoustic tags to minimize the impact on fish movement and behaviour \cite{li2024recent}, advancements in computer vision and AI to improve tracking accuracy without the need for physical tags \cite{liu2024deep}, and exploration of environmental Deoxyribonucleic Acid (DNA) or eDNA (Genetic material collected from environmental samples) methods for non-invasive population monitoring \cite{verhelst2023enhancing}.
Balancing data collection needs with fish welfare involves implementing rotation systems in tagging studies to reduce the duration of tag attachment for individual fish, using hybrid methods to reduce the reliance on any single, potentially stressful technique, and developing guidelines for the ethical use of tracking technologies in aquaculture, considering factors such as tag size relative to fish size, duration of studies, and recovery periods \cite{matley2023making, verhelst2023enhancing}. The aquaculture industry is trending towards methods that minimize stress and interference with natural behaviours while providing effective management data. This shift is driven by both ethical concerns and the recognition that improved welfare contributes to better product quality and farm productivity.}

\subsection{Real-world applications of fish tracking}
\begin{figure}[t]
    \centering
    \includegraphics[width=\linewidth]{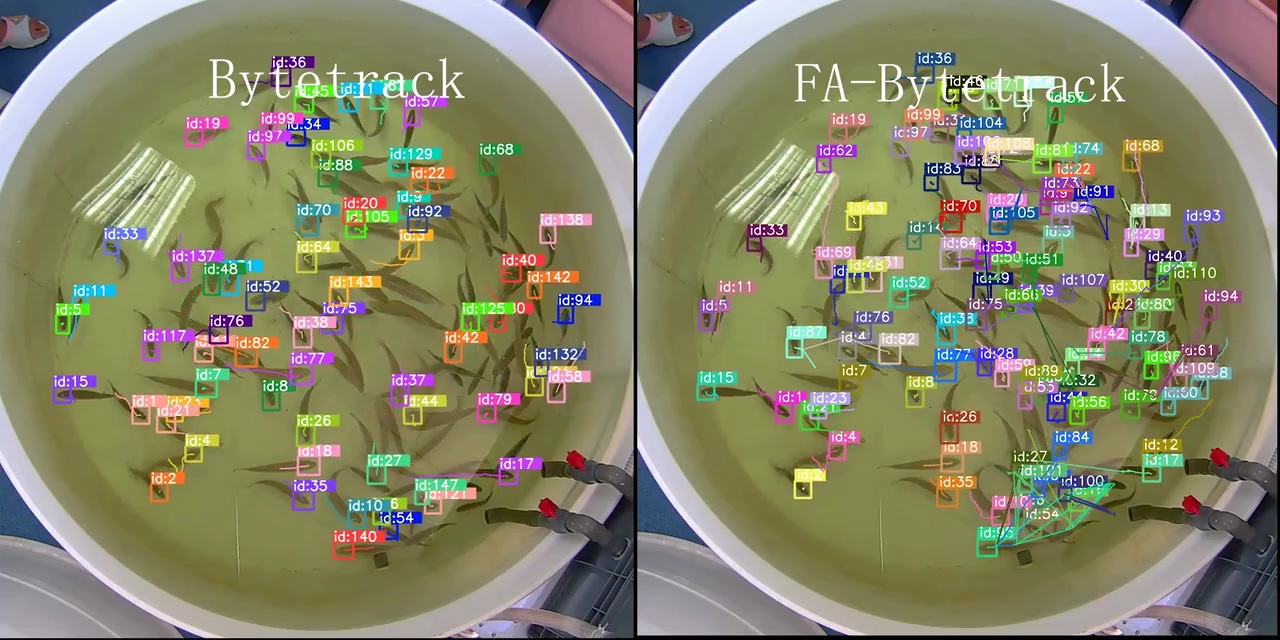}
    \caption{Comparison of ByteTrack and FA-Bytetrack in controlled
    environment (This figure
was reproduced from\cite{zhao2024fish}).}
    \label{fig:bytetrack}
\end{figure}

 \textcolor{black}{Real-world applications of fish tracking primarily span two domains: commercial aquaculture monitoring in controlled tank environments and marine ecological research in natural underwater settings. While aquaculture applications focus on fish behaviour and health monitoring through overhead cameras, underwater tracking faces more complex challenges including varying visibility, coral occlusions, and unpredictable fish movements. A recent study by Zhao et al. \cite{zhao2024fish} demonstrates the potential of advanced tracking and behaviour analysis techniques in aquaculture settings. The researchers developed a method combining an improved ByteTrack algorithm ``Fish amendments (FA)-ByteTrack'' with a spatiotemporal graph convolutional network (ST-GCN) to assess fish appetite in a controlled environment. In their experiment, they used a circular tank (1.5 m × 1.5 m × 1 m) with 120 large-mouth bass, each weighing 400-500 grams. This setup aimed to simulate the complex conditions of real aquaculture environments, including fish overlapping and occlusions. The system used a high-resolution camera (1920x1080 pixels) positioned 1.6 m above the water surface, recording at 60 frames per second. This method not only effectively solves the problems of bubble masking, intraclass variation and cross-occlusion but also achieves high-precision appetite assessment. Compared with that of ByteTrack, the average Multiple Object Tracking Accuracy (MOTA) of FA-ByteTrack increased by 21.3\%. Moreover, the tool achieved 98.47\% accuracy in appetite assessment (as shown in Fig. \ref{fig:bytetrack} ).}

 \textcolor{black}{However, it is worth noting that while this study represents a significant advancement, it was conducted in a controlled environment that does not fully reflect the challenges of real-world aquaculture settings. In commercial aquaculture operations, factors such as larger tank sizes, higher fish densities, varying water conditions, and different species could significantly impact the performance of such systems. Moreover, the study relied on a pre-trained YOLOv8 detector, which may need a large amount of high-quality labelled data for training detectors. Labelling them is time-consuming and costly.}
 
\begin{figure}[t]
    \centering
    \includegraphics[width=\linewidth]{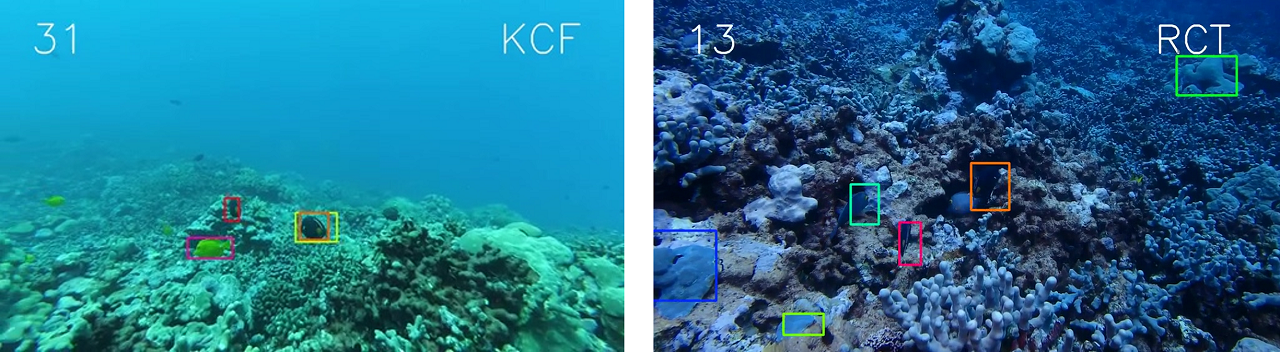}
    \caption{FISHTRAC dataset in underwater environments (This figure
was reproduced from \cite{mandel2023detection}).}
    \label{fig: FISHTRAC}
\end{figure}

 \textcolor{black}{The FISHTRAC dataset~\cite{mandel2023detection} exemplifies the challenges of real-world fish tracking in underwater environments, as shown in Fig. \ref{fig: FISHTRAC}. Unlike controlled aquarium settings, natural underwater environments present unique challenges. Fish exhibit unpredictable movements, and rapidly changing appearances, and are frequently occluded by coral or other marine life. Moreover, fish often actively attempt to evade or hide from divers, adding another layer of complexity to the tracking task. A recent work by Mandel et al. \cite{mandel2023detection} proposed a detection confidence-driven multi-object tracking to recover reliable tracks from unreliable detections. The study relied on a limited dataset of only 1800 images labelled as ``fish" from the Google Open Images Dataset \cite{kuznetsova2020open}, many of which were not from real underwater environments. The limited scale of this dataset resulted in a detector with mediocre performance, highlighting the real-world challenges of applying deep learning methods to specialized domains like underwater fish tracking.} 

 \textcolor{black}{Despite these limitations, the Robust Confidence Tracking (RCT) algorithm \cite{mandel2023detection} showed promising results.  The key strength of RCT lies in its ability to fuse low-confidence detections with a motion model and single object tracker, maintaining consistent tracks even when high-confidence detections are sparse. This capability is especially valuable in challenging underwater conditions where high-quality detections are often unavailable. The study highlights the potential of RCT for practical applications in marine biology research and ecological monitoring, particularly in scenarios with scarce labelled data and challenging detection conditions. Advanced tracking algorithms based on Bayesian filters \cite{zhao2023audio}, such as particle filters \cite{qian2019multi} and Bernoulli filters \cite{zhao2022audio}, offer additional tools for trajectory smoothing and handling outliers or mis-detection. These methods, widely used in audio-visual speaker tracking, offer the potential to be applied in underwater environments. Future research directions include developing online versions of these algorithms for edge computing, improving performance in high-density scenarios through Bayesian incorporation of appearance information, and exploring adaptive approaches that adjust tracker behaviour based on detection quality. These advancements aim to bridge the gap between controlled studies and the complex realities of commercial aquaculture, paving the way for more robust and practical fish tracking and behaviour analysis systems.}
 
\section{Fish counting}
\label{sec: counting}
\subsection{Fish counting methods based on sensor technology}

Sensor-based counting devices are usually divided into resistance counters and infrared counters. Infrared counters detect infrared signals, which are electromagnetic waves with wavelengths between 760 nm and 1 mm \cite{ewing1983infrared}. Counting based on infrared counters requires a tunnel structure to limit the movement of the fish. When a fish passes between the infrared transmitter and the receiver, the counting is completed \cite{cadieux2000intelligent, ferrero2014optical}. Although infrared sensors can count in small areas of space, their performance is affected by water depth and turbidity. At a depth of 17.9 centimetres (cm) in pure water, the intensity of the infrared light drops to 50\% \cite{santos2008monitoring}, and the presence of suspended particles can further degrade the performance of infrared counting at high turbidity levels \cite{baumgartner2012influence}. In addition to environmental factors, the accuracy of infrared sensor counting devices is susceptible to the pass rate of fish, often resulting in an underestimation of the number of fish  \cite{klapp2018ornamental, shardlow2004assessment}. This may be due to the slow swimming of some fish, the confusion when two or more fish enter the scanner unit simultaneously, and the reluctance of some fish to leave the device after entering the light tunnel, resulting in repeated scanning. Despite these limitations, infrared light can work in the dark, and the accuracy of counting can be improved by subsequent software algorithms, such as multiple object tracking (MOT) algorithms, which can solve false counting from multiple targets \cite{baumgartner2012influence, li2020nonintrusive, haas2024monitoring}.

Resistivity counters, another type of sensor-based counting method, work by detecting changes in resistance when a fish passes between two electrodes \cite{beaumont1986use, dunkley1982assessment}. Like infrared sensor counting devices, electronic resistivity counters require the fish to pass through a specific tunnel and have similar disadvantages, such as repeating counts when a fish swims multiple times in the channel and missing counts when the number of fish is large \cite{forbes2000assessment}. However, electronic resistivity counters are suitable for limited lighting and long, narrow river channels while detecting non-destructively and without requiring specific lighting conditions \cite{aprahamian1996use}.

Although both infrared and resistivity fish counters have limitations and may underestimate fish pass rates, they offer valuable tools for non-invasive fish counting in various environments. Future research should focus on developing and improving these technologies to enhance their accuracy and reliability. Potential avenues for improvement include modifying resistivity counters and exploring alternative sensors \cite{sheppard2015utility}. Researchers can provide valuable tools for effective fishery management and conservation efforts by addressing the current challenges and refining these sensor-based counting methods.

\subsection{Fish counting methods based on computer vision technology}

% \begin{figure}[t]
% \centering
% \includegraphics[width=8.5cm]{figures/counting methods.png}
% \caption{Fish counting dataset acquisition method based on machine vision.}
% \label{fig:cv_counting}
% \end{figure}

Accurate fish biomass assessment is crucial for optimizing management strategies and reducing feeding costs in the aquaculture industry \cite{cheal2020counts, pais2018effect}. Computer vision-based fish counting has gained prominence among various methods due to its non-invasive nature, low cost, and high efficiency \cite{saberioon2017application, zhang2020popu}. However, the complexity of underwater environments, including varying light conditions, backgrounds, and fish swimming patterns, poses challenges for accurate fish counting \cite{duan2015automatic, shortis2016progress, li2019detection}.

The current computer vision-based fish counting aquaculture methods focus on two main categories: image-based counting and video-based counting, as shown in Table. \ref{tab:countingmethods}. By bridging the gap between laboratory-based experiments and real-world applications, computer vision-based fish counting can become a useful tool for sustainable aquaculture management. The subsections provide more details about each category, the advancements, challenges, and future directions in the field. 

\subsubsection{Image-based fish counting method}
Image-based fish counting methods can be broadly categorized into two main approaches: detecting-based methods, which aim to detect all fish in a region, and density-based methods \cite{arteta2014interactive, fiaschi2012learning}, which estimate the number of fish by analyzing the distribution of fish schools \cite{liu2023application, saleh2023applications}.

\begin{table*}[]
\begin{threeparttable}
\caption{Different counting methods based on computer vision.}
\label{tab:countingmethods}
\begin{tabular}{@{}cccccccccc@{}}
\toprule
\begin{tabular}[c]{@{}c@{}}Study \\ Cite\end{tabular} &  \begin{tabular}[c]{@{}c@{}}Fish\\amount\end{tabular} & Dataset & Model                                                                              & Count points                                                       & \begin{tabular}[c]{@{}c@{}}Evaluation \\ index\end{tabular}              & Results                                                         & Advantages                                                                                            & Limitations                                                                                                                       & References                \\ \midrule
Tank                                                  & 100    & 786     & $\mathrm{MAN}$\tnote{1}                                                                   & Center                                                             & Accuracy                                                                 & $97.12 \%$                                                      & \begin{tabular}[c]{@{}c@{}}Better \\ generalization \\ ability\end{tabular}                           & \begin{tabular}[c]{@{}c@{}}larger error for \\ areas with high \\ fish density\end{tabular}                                       & \cite{yu2022counting}            \\
Tank                                                  & -      & 4000    & DG-LR\tnote{2}                                                                       & \begin{tabular}[c]{@{}c@{}}Fish- \\ Connected \\ Area\end{tabular} & $\mathrm{R}^2$\tnote{3}                                                           & $96.07 \%$                                                      & \begin{tabular}[c]{@{}c@{}}No need to detect \\ every fish\end{tabular}                               & \begin{tabular}[c]{@{}c@{}}No complex \\ environments\end{tabular}                                                                & \cite{zhang2020automatic}        \\
\begin{tabular}[c]{@{}c@{}}Net \\ Cage\end{tabular}   & 214    & 1501    & \begin{tabular}[c]{@{}c@{}}Hybrid Neural \\ Network\end{tabular}                   & Center points                                                      & Accuracy                                                                 & $95.06 \%$                                                      & \begin{tabular}[c]{@{}c@{}}Improves model \\ performance \\ without losing \\ resolution\end{tabular} & \begin{tabular}[c]{@{}c@{}}Does not \\ describe the \\ distribution of \\ fish school \\ gathering and \\ dispersing\end{tabular} & \cite{zhang2020automatic1}       \\
Cage                                                  & 62     & 200     & $\mathrm{RCNN}$\tnote{4}                                                                    & Bounding Box                                                       & Accuracy                                                                 & $92.4 \%$                                                       & \begin{tabular}[c]{@{}c@{}}Reduce count \\ errors due to \\ repeating \\ detections\end{tabular}      & \begin{tabular}[c]{@{}c@{}}Repeating \\ detection and \\ wrong detection \\ in high contrast \\ areas\end{tabular}                & \cite{xu2020detection}           \\
Counter                                               & 1000   & 1500    & \begin{tabular}[c]{@{}c@{}}Background \\ Subtraction \\ Kalman filter\end{tabular} & Blob                                                               & \begin{tabular}[c]{@{}c@{}}Average \\ precision\end{tabular}             & $97.47 \%$                                                      & \begin{tabular}[c]{@{}c@{}}Automatic \\ counting, low \\ cost\end{tabular}                            & \begin{tabular}[c]{@{}c@{}}No detailed \\ analysis of the \\ number of fish \\ in the system \\ per unit time\end{tabular}        & \cite{albuquerque2019automatic}  \\
Containers                                            & 600    & 4000    & $\mathrm{CNN}$                                                                     & Contours                                                           & Accuracy                                                                 & $99.17 \%$                                                      & \begin{tabular}[c]{@{}c@{}}Threshold adapts \\ to different \\ numbers of fish\end{tabular}           & \begin{tabular}[c]{@{}c@{}}Pure white \\ background, no \\ noise\end{tabular}                                                     & \cite{lainez2019automated}       \\
Dishpan                                               & 100    & -       & \begin{tabular}[c]{@{}c@{}}Local \\ Normalization \\ Filter\end{tabular}           & Pixel Area                                                         & \begin{tabular}[c]{@{}c@{}}Accuracy \\ F-measure\end{tabular}            & \begin{tabular}[c]{@{}c@{}}$99.8 \%$ \\ $98.83 \%$\end{tabular} & \begin{tabular}[c]{@{}c@{}}automated \\ system.\end{tabular}                                          & \begin{tabular}[c]{@{}c@{}}Small sample \\ size\end{tabular}                                                                      & \cite{coronel2018identification} \\
Aquarium                                              & 350    & 1000    & \begin{tabular}[c]{@{}c@{}}Background \\ Subtraction\end{tabular}                  & Contours                                                           & Accuracy                                                                 & $95.57 \%$                                                      & \begin{tabular}[c]{@{}c@{}}Portable, low \\ cost\end{tabular}                                         & \begin{tabular}[c]{@{}c@{}}Need a fixed \\ size of fish and \\ a certain area\end{tabular}                                        & \cite{hernandez2018development}  \\
% Aquarium                                              & 9      & -       & \begin{tabular}[c]{@{}c@{}}Adaptive \\ Thresholding\end{tabular}                   & Skeleton                                                           & \begin{tabular}[c]{@{}c@{}}Average \\ counting error\end{tabular}        & $6 \%$                                                          & \begin{tabular}[c]{@{}c@{}}Solve the \\ overlapped-fish \\ problem cleverly\end{tabular}              & \begin{tabular}[c]{@{}c@{}}Only adapted to \\ relatively small \\ fish densities\end{tabular}                                     & \cite{le2017automated}           \\
Net Cage                                              & 250    & 1000    & $\mathrm{PTV}$\tnote{5}                                                                   & Centroid                                                           & Detection rate                                                           & $90 \%$                                                         & \begin{tabular}[c]{@{}c@{}}Potential \\ application or \\ industrial \\ aquaculture\end{tabular}      & \begin{tabular}[c]{@{}c@{}}Affected by \\ background \\ noise sensitivity\end{tabular}                                            & \cite{abe2017many}               \\
Aquarium                                              & 100    & 600     & LS-SVM\tnote{6}                                                         & Skeleton                                                           & Accuracy                                                                 & $98.73 \%$                                                      & \begin{tabular}[c]{@{}c@{}}Good \\ generalization\end{tabular}                                        & \begin{tabular}[c]{@{}c@{}}Assume that the \\ size of fish is \\ similar\end{tabular}                                             & \cite{fan2013automate}           \\
Aquarium                                              & 300    & 3200    & MSENet\tnote{7}                                                                             & Centroid                                                           & MAE\tnote{8}                                                                      & 3.33                                                            & \begin{tabular}[c]{@{}c@{}}Lightweight and \\ low \\ computation \\ costs\end{tabular}                & \begin{tabular}[c]{@{}c@{}}limited to a \\ scene with a \\ fixed \\ viewpoint\end{tabular}                                        & \cite{li2023lightweight}         \\
Long Channel                                          & 300    & 1318    & \textbf{YOLOv5-Nano}                                                                        & Bounding Box                                                       & \begin{tabular}[c]{@{}c@{}}Average \\ precision\end{tabular} & $96.4 \%$                                                       & \begin{tabular}[c]{@{}c@{}}Solves the \\ problem of \\ missing fish fry\end{tabular}                  & \begin{tabular}[c]{@{}c@{}}Occlusion still \\ causes some \\ fish to be \\ incorrectly \\ detected\end{tabular}                   & \cite{zhang2023dynamic}          \\ \bottomrule
\end{tabular}
        \begin{tablenotes}
            \item[1] MAN (Multi-modules and attention mechanism)
            \item[2] DG-LR (Image density grading and local regression)
            \item[3] R2 (Coefficient of determination)
            \item[4] RCNN (Region-based convolutional neural network)
            \item[5] PTV (Particle tracking velocimetry)
            \item[6] LS-SVM (Least squares support vector machine)
            \item[7] MSENet (A lightweight network based on SENet )
            \item[8] MAE (Mean absolute error )
        \end{tablenotes}
\end{threeparttable}
\end{table*}

 Early studies focused on detecting-based methods, which rely heavily on the accuracy of fish image segmentation from the background \cite{tran2018determination}. These methods, such as artificial neural networks (BPNN) \cite{newbury1995automatic}, showed potential for automatic fish counting in scenarios with a limited number of fish. However, they often struggled with complex adhesions in fish images and overlapping fish \cite{al2018automatic, hernandez2018development}.
To address the challenges of overlapping fish, adaptive segmentation algorithms were developed to extract the geometric features of fish \cite{fan2013automate}. Combined with machine learning models like LS-SVM, these algorithms showed improved counting accuracy compared to BPNN models, particularly in scenarios with similar fish sizes and low stocking densities. However, the performance of these models degrades when faced with high fish densities and changing geometric shapes due to fish overlap \cite{aliyu2020incorporating}.
Further advancements in fish image segmentation were made by introducing more general adaptive thresholding methods and skeleton extraction-based methods to handle overlapping fish \cite{le2017automated}. While these methods performed well under controlled laboratory conditions, their accuracy diminished in real-world aquaculture environments, where factors such as high fish school density, poor visibility, and insufficient light posed significant challenges \cite{abe2017many}.

Efforts to mitigate issues related to light, noise, and feature recognition led to the development of segmentation methods that combined local normalized filters and iterative selection thresholds \cite{coronel2018identification}. Although these methods demonstrated high performance in correcting non-uniform lighting, reducing noise, and identifying features, the unique challenges posed by aquaculture settings, such as fish shadows caused by water refraction and continuous movement of shoals, continued to affect segmentation accuracy and limit the effectiveness of traditional computer vision methods for fish counting \cite{hernandez2018development}.

Introducing deep learning techniques has opened new avenues for fish counting in aquaculture. With the increasing availability of fish datasets, deep learning models have been applied to this domain, offering strong adaptability and easy transformation without requiring complex feature extraction work \cite{labao2019cascaded, xu2020detection}. Convolutional neural networks (CNNs) have achieved high accuracy in detecting and counting fish of different sizes by adjusting different thresholds \cite{lainez2019automated}.

Density-based methods, which estimate the number of fish by mapping input images to corresponding density maps, have also shown promise in fish counting applications. These methods provide additional information about the spatial distribution of fish, which can be valuable for various purposes \cite{liu2019adcrowdnet}. Hybrid neural network models, such as those combining MCNN and DCNN architectures, have been proposed to improve fish counting accuracy, outperforming traditional CNNs and MCNNs \cite{zhang2020popu, yu2022counting}.

Despite the advancements made in fish counting methods, several challenges remain. Density-based methods are sensitive to the degree of occlusion, with higher fish densities leading to greater errors. Moreover, variations in water quality, light conditions, camera angle, water depth, and surface refraction can cause significant differences in the appearance of fish across different farming environments, affecting the accuracy and generalization ability of the counting models.
To address these challenges, future research should create more comprehensive and diverse datasets that capture the variability encountered in real-world aquaculture settings. Efforts should also be directed towards improving counting accuracy, model generalization ability in high-density areas, and maintaining accuracy under different pond conditions.

\subsubsection{Video-based fish counting method}

Video-based counting methods offer a more comprehensive approach to enumerating fish than static image-based techniques, leveraging temporal information and movement patterns \cite{shafait2016fish}. These methods can be broadly categorized into frame-by-frame analysis, tracking-based counting, and segmentation-based counting \cite{li2021automatic}. 

\textcolor{black}{The frame-by-frame analysis based methods extend the image-based techniques to video sequences, applying detection and counting algorithms to each frame. Zhao et al. \cite{zhao2019algorithm} proposed a method using background subtraction and object detection to count fish in video streams, achieving higher accuracy than single-image based counting methods due to the ability to average counts over multiple frames. Like this work, Pai et al. \cite{pai2022computer} used the YOLOv5 model \cite{tang2024hic} to detect and count fish in video streams and get similar results. This approach is relatively simple to implement and can leverage existing image-based algorithms. However, it does not fully use temporal information, may struggle with rapid fish movements between frames, and is prone to double-counting or missing fish due to occlusions.}

\begin{figure}[t]
    \centering
    \includegraphics[width=\linewidth]{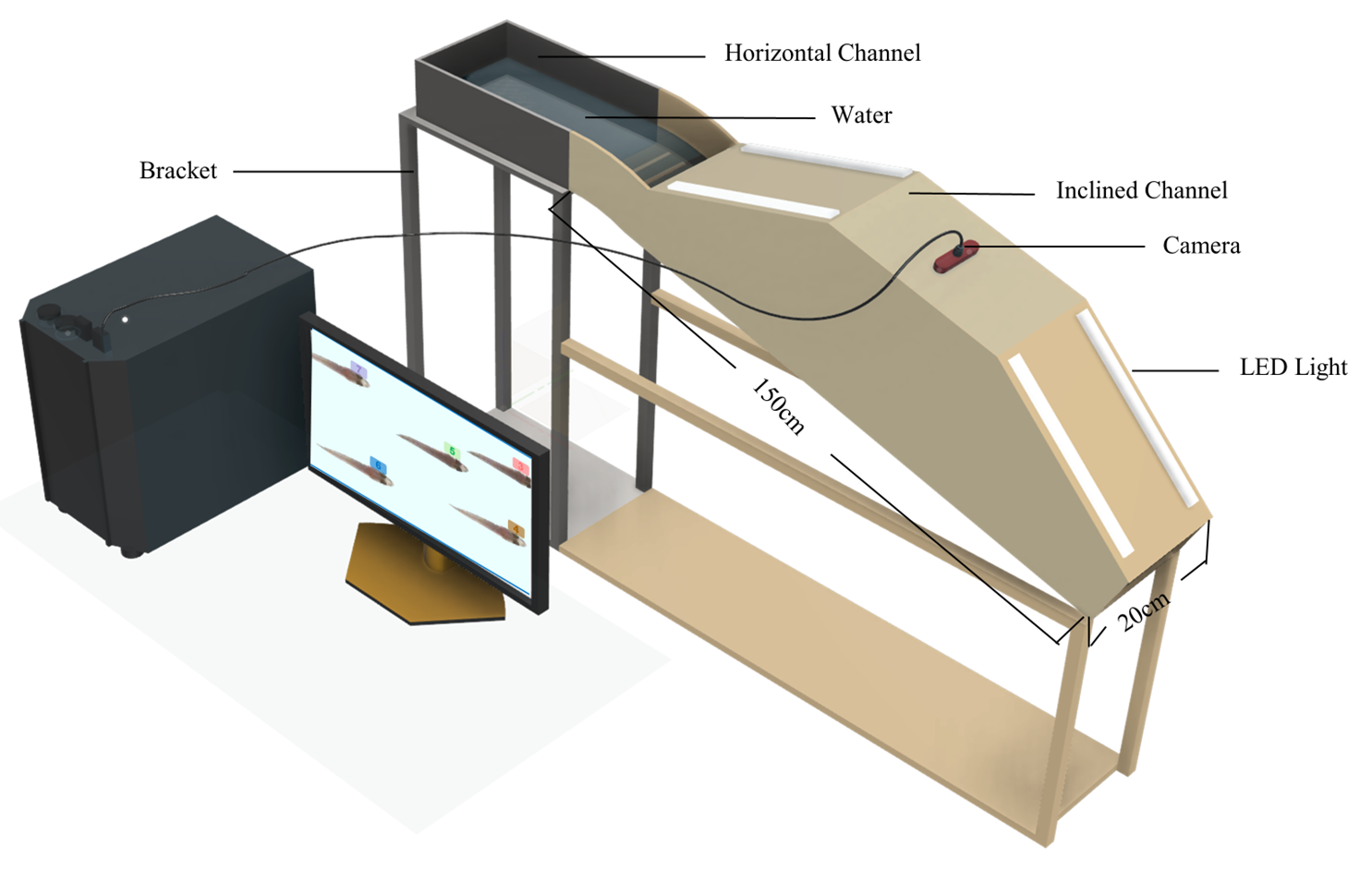}
    \caption{Overview of the fry counting system. Fish is poured into a channel with water flow, LED lights provide stable illumination. An industrial camera records the fry swimming process and transmits it to a computer (or edge device), and the device counts the number of fry passed in real-time through the designed method.}
    \label{fig:counting_machine}
\end{figure}

\textcolor{black}{Tracking-based counting methods follow individual fish across frames, offering insights into fish movement patterns and behaviours. In addition, the tracking-based counting methods can deal with the temporary occlusions, and reduce double-counting errors \cite{liu2021multi}. Albuquerque et al. \cite{albuquerque2019automatic} and Zhou et al. \cite{zhou2022kinematic} developed a system combining object detection with multiple object tracking to count fish in dynamic videos for large-scale counting in practical production. This method uses segmentation and association, and can effectively solve the problem of mis-detection caused by adhesive fingerlings in mechanical counters (as shown in Fig. \ref{fig:counting_machine}). However, in practical applications, these algorithms often struggle to achieve a balance between counting accuracy and processing speed, limiting their applicability in real-time commercial settings. Addressing this challenge, Zhang et al. \cite{zhang2023dynamic} proposed a novel combination of the SORT algorithm \cite{zhao2019algorithm} and YOLOv5-Nano \cite{ghahremannezhad2023object} for fry tracking and counting. This innovative approach achieves accurate dynamic counting with even faster running speeds, representing a significant step towards real-time fish counting in aquaculture environments. Integrating lightweight deep learning models with efficient tracking algorithms showcases a promising direction for future research, potentially enabling more widespread adoption of video-based counting methods in commercial aquaculture operations.}

\textcolor{black}{Segmentation-based counting methods focus on isolating fish from the background in each frame and then counting the segmented objects. Instance segmentation methods have emerged as a promising solution for effectively segmenting individual fish in dense scenes, and combining these methods with advanced detectors has the potential to achieve real-time performance \cite{liu2021high}. Wang et al. \cite{wang2024automated} propose a lightweight instance segmentation model based on YOLOv8 \cite{sohan2024review}, realizing high-density and multi-target fish segmentation and counting in an occluded environment. This approach demonstrates the potential of segmentation-based methods in handling complex aquaculture scenarios. However, the segmentation performance is notably diminished in scenarios of significant occlusions, presenting a significant challenge for dense fish populations. To address this limitation, recent advancements in computer vision, such as the Segment Anything Model (SAM) \cite{kirillov2023segment}, offer new possibilities. The SAM model is pre-trained on a large-scale dataset and exhibits exceptional robustness in segmentation tasks across various domains. Although no literature is found for fish counting based on SAM, its excellent segmentation capabilities hold promise for future applications in aquaculture. Looking ahead, integrating the robust segmentation abilities of SAM with specialized fish counting algorithms could potentially overcome the limitations of current methods, particularly in highly occluded environments. It could be a promising direction to adapt SAM for fish-specific segmentation tasks and optimise its real-time performance to meet the demands of commercial aquaculture operations.}

\subsection{Fish counting methods based on acoustic technology}
Acoustic technology for fish counting can be divided into two main categories: acoustic imaging and hydroacoustic methods. While underwater visible imaging suffers from limitations due to light attenuation caused by water absorption and scattering, resulting in blurred images and reduced image quantity as shooting distance increases, acoustic-based counting methods offer a viable alternative. Sound waves can travel far through water without significant attenuation, making them suitable for situations where visual counting is inappropriate or ineffective.

\subsubsection{Acoustic imaging methods}
Multi-beam imaging sonar such as Dual-frequency Identification Sonar (DIDSON) and Adaptive Resolution Imaging Sonar (ARIS) are normally used to monitor migratory fish in rivers \cite{martignac2015use}. These systems produce high-resolution underwater sonar video output without the need for underwater light, allowing for fish counting and measuring directly from the footage, even in turbid waters and overnight \cite{lagasse2017assessment}.

\begin{figure}[t]
    \centering
    \includegraphics[width=\linewidth]{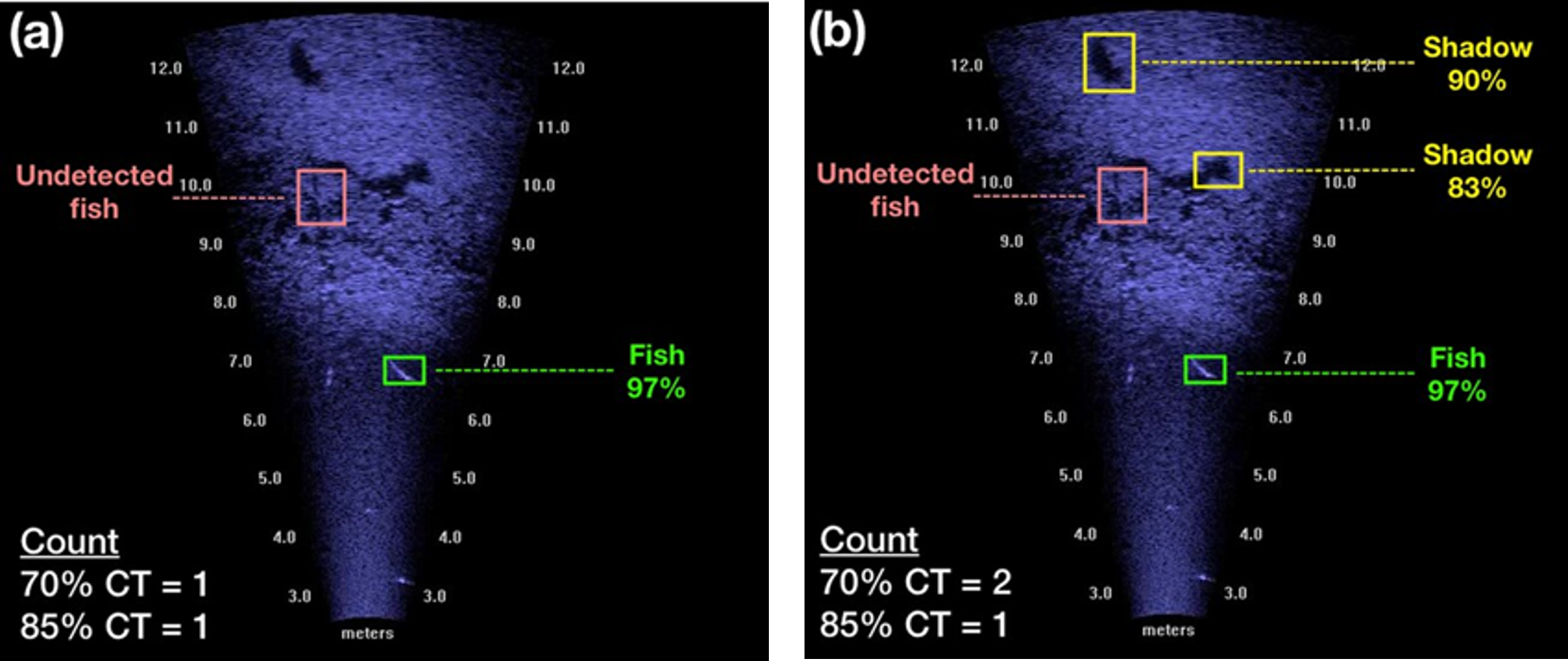}
    \caption{Example of DIDSON image counting (This figure was reproduced from \cite{connolly2023out} with minor changes).}
    \label{fig: sona_counting}
\end{figure}

DIDSON \cite{belcher2002dual} is a multi-beam sonar system frequently used to acquire underwater acoustic images for fish identification and counting (as shown in Fig. \ref{fig: sona_counting}). As DIDSON uses sound instead of light, it is not affected by water turbidity and can collect data during both day and night \cite{cronkite2006use, maxwell2004feasibility}. However, studies have shown that manual counting of DIDSON data can be time-consuming and prone to errors, with large deviations between operators \cite{lagarde2020situ, maxwell2007assessing}. This may be because Echoview repeatedly calculated at nearly stationary horizontal positions within the DIDSON field of view \cite{petreman2014observer}.

To reduce the time and cost of DIDSON data processing, various subsampling methods can be employed, with automation-assisted subsampling being the best method to reduce the cost of estimating migratory fish populations in rivers \cite{eggleston2020improved}. 
Multi-beam echogram processing software, such as Echoview or DIDSON Control and Display software, can partially perform fish detection and counting functions \cite{boswell2008semiautomated, hughes2015combining}. Echoview uses a Component Object Model (COM) interface that allows users to build customized pre-processing and post-processing scripting modules, streamlining the processing method and providing the ability to refine fish counting using various fish detection parameters \cite{berghuis2008performance, han2009didson}. However, the echograms of the video-like data files generated by DIDSON require manual counting, which is tedious, time-consuming, and can produce large errors for large datasets \cite{mora2015estimating}. Semi-automatic post-processing of imaging sonar data is possible using existing software (e.g., Echoview Software Pty Ltd., Hobart, Australia) \cite{boswell2008semiautomated, kang2011semiautomated}. but the process still requires manual calibration for non-fish target noise, which is cumbersome and inefficient. Furthermore, post-processing software can be very expensive, limiting its accessibility for many researchers and practitioners.

Digital image-processing technology offers an inexpensive and rapid alternative that has been successfully applied in various scientific fields. Several studies have focused on the automatic processing of fish targets in imaging sonar data. For example, K-nearest neighbour background subtraction with DeepSort target tracking to track and count fish automatically \cite{shen2024identification} and GPNet, a novel encoder-decoder network with global attention and point supervision, to boost sonar image-based fish counting accuracy \cite{duan2023boosting}.

The new generation of acoustic cameras includes ARIS (Sound Metrics Corp, WA, USA), which operates at higher frequencies compared to DIDSON, offering greater flexibility and improved image resolution \cite{jones2021adaptive, shahrestani2017detecting}.  A comparison of fish monitoring data based on the ARIS sonar system and the GoPro camera showed that the detection rate of the sonar-based system was 62.6\% (compared to the amount captured by the net), exceeding the 45.4\% of the camera-based system \cite{egg2018comparison}.

While sonar imaging counting methods are powerful tools for gathering fish abundance estimates in difficult-to-observe, structurally complex, chaotic, and dark environments, they can still be disturbed by various types of underwater noise. Additionally, sonar imaging equipment is relatively expensive and requires professional personnel to conduct analysis, making it more suitable for investigating fish abundance in ocean fishing and river ports \cite{holmes2006accuracy, jing2017method}.

These recent advancements in digital image-processing techniques showcase the growing interest in developing efficient and accurate methods for automatic fish tracking and counting in sonar data. By leveraging the power of deep learning and computer vision algorithms, these approaches aim to overcome the limitations of manual processing and provide more reliable and scalable solutions for aquaculture monitoring and management.
However, while these methods show promising results, they still face challenges such as dealing with occlusions, varying fish densities, and the need for large annotated datasets for training. Future research should address these limitations and develop more robust and generalizable algorithms that can be easily adapted to different sonar imaging systems and underwater environments.

Table \ref{tab: acoustic imaging} summarizes the advantages and disadvantages of sonar imaging counting methods in aquaculture and their practical applications. Despite the limitations, acoustic counting methods remain valuable for monitoring fish populations in challenging underwater environments where visual counting methods may be impractical or ineffective.

\begin{table*}[]
\begin{threeparttable}
\caption{A comparison of different methods based on acoustic imaging.}
\label{tab: acoustic imaging}
\centering
\begin{tabular}{@{}ccccccccc@{}}
\toprule
Site      & Technology & Software                                                                                 & $\mathrm{MHz}$                                                          & Metrics                                                                                        & Results                                                                            & Advantages                                                                                                      & Limitations                                                                                                            & References                                                            \\ \midrule
River     & ARIS       & Echoview                                                                                 & \begin{tabular}[c]{@{}c@{}}$1.1 \mathrm{MHz}$ \\ frequency\end{tabular} & Accuracy                                                                                       & $84 \%$                                                                            & \begin{tabular}[c]{@{}c@{}}Distinguishes \\ downstream \\ moving fish from \\ other objects\end{tabular}        & \begin{tabular}[c]{@{}c@{}}Results vary \\ among operators\end{tabular}                                                & \cite{helminen2021object}                                                    \\
Lagoon    & ARIS       & Sound Metrics                                                                            & $1.8 \mathrm{MHz}$                                                      & $R^2$                                                                                          & 0.99                                                                               & \begin{tabular}[c]{@{}c@{}}Consistent results \\ with manual\end{tabular}                                       & \begin{tabular}[c]{@{}c@{}}The results varied \\ greatly among \\ different \\ operators\end{tabular}                  & \cite{lagarde2020situ}                                                       \\
River     & ARIS       & \begin{tabular}[c]{@{}c@{}}ARIS \\ software Fish\end{tabular}                            & $1.8 \mathrm{MHz}$                                                      & F1-scores                                                                                      & $75 \%$                                                                            & \begin{tabular}[c]{@{}c@{}}Faster and no \\ post-processing\end{tabular}                                        & \begin{tabular}[c]{@{}c@{}}Underestimates \\ total fish count\end{tabular}                                             & \cite{le2023automatic}                                                       \\
Reservoir & ARIS       & \begin{tabular}[c]{@{}c@{}}KNN \\ background \\ subtraction and \\ DeepSort\end{tabular} & $1.8 \mathrm{MHz}$                                                      & Accuracy                                                                                       & $73 \%$                                                                            & \begin{tabular}[c]{@{}c@{}}Automatic \\ calibration saves \\ data processing \\ time\end{tabular}               & \begin{tabular}[c]{@{}c@{}}Unable to \\ identify fish in \\ bottom \\ background, long \\ processing time\end{tabular} & \cite{shen2024identification}                                                \\
River     & ARIS       & ARISfish                                                                                 & $3.0 \mathrm{MHz}$                                                      & Detection Rate                                                                                 & $62.6 \%$                                                                          & \begin{tabular}[c]{@{}c@{}}Counts fish $>100$ \\ mm in night and \\ turbid conditions\end{tabular}              & \begin{tabular}[c]{@{}c@{}}May not detect \\ small fish\end{tabular}                                                   & \cite{egg2018comparison}                                                     \\
River     & DIDSON     & Echoview 6.0                                                                             & $1.8 \mathrm{MHz}$                                                      & Accuracy                                                                                       & $83.7 \%$                                                                          & \begin{tabular}[c]{@{}c@{}}Avoids manual \\ counting errors \\ and biases\end{tabular}                          & \begin{tabular}[c]{@{}c@{}}Time-consuming \\ calculations\end{tabular}                                                 & \cite{eggleston2020improved}                                                 \\
River     & DIDSON     & Sound Metrics                                                                            & $1.8 \mathrm{MHz}$                                                      & F1 scores                                                                                      & $79 \%$                                                                            & \begin{tabular}[c]{@{}c@{}}Performs well \\ using direct, \\ shadow, and \\ combined \\ detections\end{tabular} & \begin{tabular}[c]{@{}c@{}}Low fish \\ densities in each \\ image\end{tabular}                                         & \cite{connolly2023out}                                                       \\
Reservoir & DIDSON     & \begin{tabular}[c]{@{}c@{}}NN-EKF2; \\ Echoview\end{tabular}                             & $1.8 \mathrm{MHz}$                                                      & \begin{tabular}[c]{@{}c@{}}Error compared \\ with the manual \\ detection results\end{tabular} & Less than $5 \%$                                                                   & \begin{tabular}[c]{@{}c@{}}Less calculation \\ and easy to \\ implement\end{tabular}                            & \begin{tabular}[c]{@{}c@{}}Inaccurate when \\ targets overlap\end{tabular}                                             & \cite{jing2017method}                                                        \\
River     & DIDSON     & \begin{tabular}[c]{@{}c@{}}Sound Metrics; \\ Echoview\end{tabular}                       & $1.2 \mathrm{MHz}$                                                      & Accuracy                                                                                       & \begin{tabular}[c]{@{}c@{}}$90 \%$ (upstream) \\ $41 \%$ (downstream)\end{tabular} & \begin{tabular}[c]{@{}c@{}}Estimates \\ potamodromous \\ fish passage in \\ large lakes\end{tabular}            & \begin{tabular}[c]{@{}c@{}}High processing \\ times and costs\end{tabular}                                             & \cite{petreman2014observer}                                                  \\
River     & DIDSON     & \begin{tabular}[c]{@{}c@{}}Manual \\ counting\end{tabular}                               & $1.8 \mathrm{MHz}$                                                      & \begin{tabular}[c]{@{}c@{}}Average Percent \\ Error (APE)\end{tabular}                         & $5.4 \%$                                                                           & \begin{tabular}[c]{@{}c@{}}Not limited by \\ surface \\ disturbances or \\ turbidity\end{tabular}               & \begin{tabular}[c]{@{}c@{}}Shadowing from \\ passing fish\end{tabular}                                                 & \begin{tabular}[c]{@{}c@{}}\cite{maxwell2007assessing}\end{tabular} \\
River     & DIDSON     & Hand-counter                                                                             & $1.8 \mathrm{MHz}$                                                      & \begin{tabular}[c]{@{}c@{}}Coefficient of \\ Variation (CV)\end{tabular}                       & $9.63 \%$                                                                          & \begin{tabular}[c]{@{}c@{}}Better acoustic \\ target \\ identification and \\ resolution\end{tabular}           & \begin{tabular}[c]{@{}c@{}}Data loss on \\ small fish in \\ highly turbulent \\ environments\end{tabular}              & \cite{holmes2006accuracy}                                                 \\ \bottomrule
\end{tabular}

\end{threeparttable}
\end{table*}

\subsubsection{Hydroacoustic methods}
Acoustic echo-sounding is one of the most popular methods for estimating fish abundance due to their simplicity and non-invasive nature \cite{wanzenbock2003quality}. These methods rely on the physical characteristics of the target and the water medium. When an echo sounder's transducer emits an acoustic wave, it spreads through the water and encounters the target object. Due to the difference in acoustic impedance between the object and the water medium, the object scatters the incident acoustic wave, and a portion of it is backscattered to the transducer, known as the echo signal \cite{mesiar1990development, zhao2003estimation}.

The target's depth can be measured according to the interval between the acoustic emission and the reception of the target's echo. By analyzing the strength and structure of the echo signal, the intensity, number, and distribution of the target can be estimated. The Echo Integration method is one of the main methods for underwater acoustic assessment of fish stocks. It calculates the number of fish by dividing the integral value of the echo intensity of fish in the sampling unit area by the ultrasonic reflectance of individual fish (target intensity, TS). Several studies have used the echo integration technique to estimate the number of fish based on the backscattering echoes observed with an echo sounder \cite{nishimori2009development, takao1996dual}.

Although the sound intensity reflected by a shoal is related to the number of fish \cite{simmonds2008fisheries}, the use of echo sounders in fish tanks and cages presents several challenges \cite{espinosa1994acoustical}. Reverberation in a cage can occur due to the echo of an acoustic signal from the boundary, necessitating the removal of the cage boundary signal during counting \cite{conti2006acoustical}. Another issue with acoustic estimation of fish populations is shadow utility, which is needed to compensate for the attenuation of echo strength when dense shoals are in focus \cite{zhao2003estimation}. To investigate the possibility of using commercial echo sounders for real-time fish counting in offshore cages, a study by \cite{sthapit2020method} employed an echosounder and echo-integration technology. The experimental results showed that the proposed method could achieve more than 90\% estimation accuracy \cite{sthapit2019algorithm}, indicating its reliability for future fish management decisions.

Despite the increasing use of underwater echo sounders in fishery research, their application is subject to interference from various factors, such as differences in instrument performance, the blind area of the echo sounder itself, external environmental factors, and the evasive behaviour of fish in response to survey ships and sound waves \cite{bjordal2020hydroacoustic, godlewska2009hydroacoustic}. Furthermore, echosounders are expensive and technically demanding, making them unsuitable for factory aquaculture needs. Future research should focus on reducing the cost of instruments or developing alternative instruments suitable for promotion to meet the actual needs of aquaculture.

\subsection{Comparative analysis and trends in fish counting methods}
\textcolor{black}{Fish counting methods in aquaculture can be broadly categorized into sensor-based, computer vision-based, and acoustic-based approaches, each with distinct advantages and limitations.
Sensor-based counting, including infrared and resistivity counters, offers real-time data but may be limited by water conditions and fish density \cite{li2021automatic}. Infrared counters are non-invasive but struggle in turbid waters, while resistivity counters can operate in various conditions but may require fish to pass through specific channels \cite{li2020nonintrusive, yang2021computer}.
Computer vision-based counting is divided into image-based and video-based methods. Image-based counting is cost-effective and can handle large datasets but may struggle with overlapping fish \cite{albuquerque2019automatic}. Video-based counting offers more dynamic analysis, and better handling of movement and overlaps, but requires more computational power \cite{pai2022computer, zhou2022kinematic}.
Acoustic-based counting excels in challenging environments, including acoustic imaging and hydroacoustic methods \cite{li2024recent}. Acoustic imaging (e.g., DIDSON) works well in turbid waters and at night but can be costly \cite{maxwell2004feasibility, petreman2014observer}. Hydroacoustic methods are effective for quantifying fish density and biomass at large scales but may lack species-specific accuracy \cite{mesiar1990development, conti2006acoustical}.}

\textcolor{black}{Analysis of research literature reveals that the selection of fish tracking methods is closely tied to aquaculture environments and infrastructure types. In intensive pond and tank systems, computer vision-based methods have been widely studied and implemented \cite{aquino2020trend}, demonstrating particular success in controlled environments. Research in large-scale marine farming has explored various approaches, including acoustic methods for deep-water applications \cite{le2023automatic}. Recent studies have increasingly focused on AI-integrated tracking systems \cite{eggleston2020improved}, while research in open-water aquaculture settings has predominantly investigated acoustic tracking methods due to their effectiveness in larger volumes \cite{heenan2017long}.
The adoption trends are shaped by environmental conditions, species being counted, regulatory requirements, and economic factors. Due to cost constraints, developing regions often rely on manual counting or basic sensor technology.
Welfare considerations are increasingly important in counting method selection. Noninvasive methods like computer vision and acoustic imaging are preferred to minimize fish stress. Ongoing research focuses on improving accuracy while reducing the impact on fish behaviour.}

\subsection{Real-world applications in fish counting}
\begin{figure}[t]
    \centering
    \includegraphics[width=\linewidth]{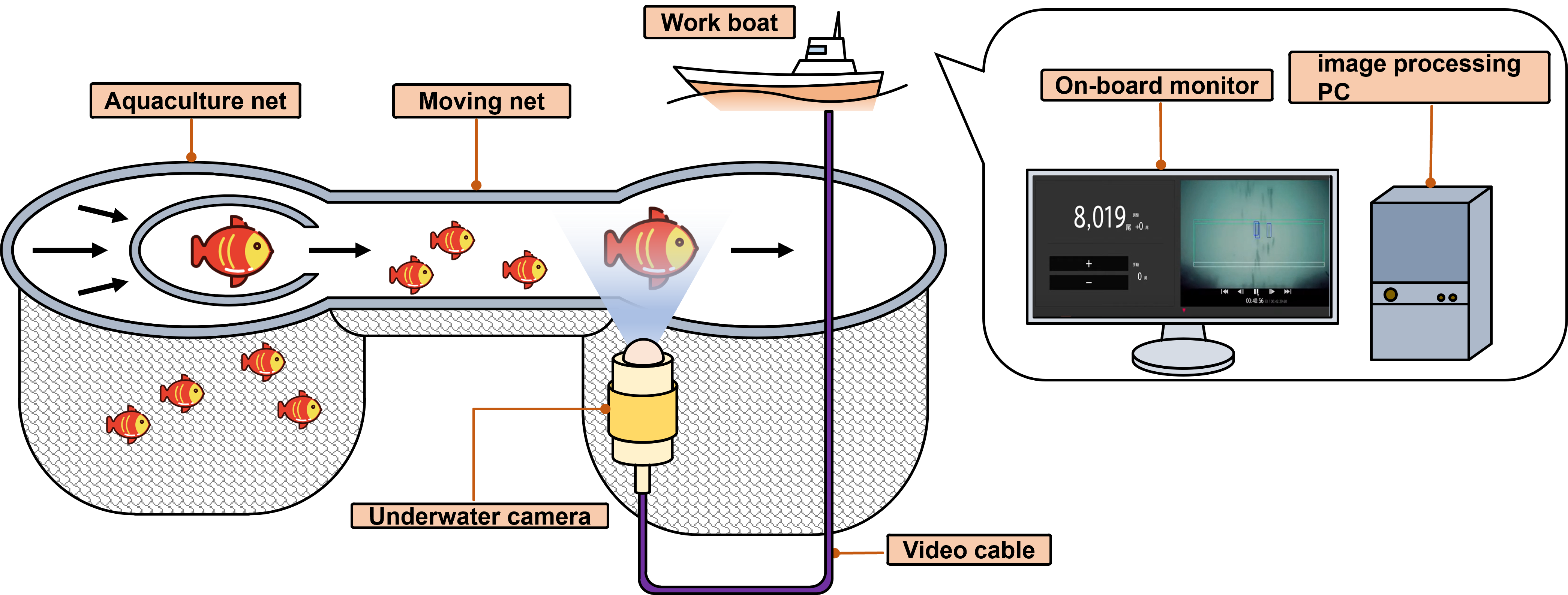}
    \caption{Example of the counting method developed by YMS.}
    \label{fig:real_counting}
\end{figure}

\textcolor{black}{Fish counting is a critical aspect of aquaculture management, essential for controlling production volume, feed usage, and regulatory reporting. Yanmar Marine Systems Co., Ltd. (YMS) developed a real-world application (as shown in Fig. \ref{fig:real_counting}) of automated fish counting to address the challenges faced in tuna farming \footnote{\url{https://www.yanmar.com/global/about/technology/vision2/fish_counting_system.html}}. Traditionally, farm fish counting relied on labour-intensive and potentially inaccurate methods such as visual counting from camera footage or subtractive counting of dead fish. To overcome these limitations, YMS developed an automated fish counting system aimed at achieving 98\% or higher accuracy in real-time tuna counting. This system integrates advanced image recognition technologies and consists of underwater cameras, a dedicated image processing computer, and specialized software. Key features include real-time counting using simultaneous imaging and count recognition, detection of environmental disturbances, and remote adjustment for optimal imaging conditions. The system is beneficial when introducing young Pacific bluefin tuna into nets, during the division of fish into multiple nets during growth phases, and for daily inventory management and feed usage monitoring. In testing, it achieved the target of 98\% or higher accuracy, leading to its commercial release in December 2020. This automated system significantly reduces the workload on farm workers while maintaining high accuracy, demonstrating the practical application of computer vision and image processing techniques in commercial aquaculture settings and potentially revolutionizing farm management practices.}

\textcolor{black}{Another typical example is a large commercial tilapia hatchery that implemented an advanced dynamic fry counting system using 32 cm wide channels (as shown in Fig. \ref{fig:counting_machine}), high-resolution cameras, and a central server running YOLOv5-Nano \cite{ghahremannezhad2023object} detection model and an improved SORT algorithm \cite{zhang2023dynamic}. The system uses several innovative approaches to address common challenges such as occlusion, high density, and adhered fry. The improved SORT algorithm maintained consistent tracking of individual fry across frames, even in high-density situations. The detection model was trained on diverse datasets including edge cases to handle adhered fry. The optimized channel design, with careful calibration of width and inclination, encouraged fry separation and reduced occlusion. Temporal information from multiple frames was utilized to resolve ambiguities caused by occlusion or high density. After six months, the incubator reported significant improvements: fry count accuracy increased to within 3-4\% (from 20\% variance), counting time was reduced by 85\%, overfeeding decreased by 12\%, and fry survival rates improved by 3\%. The implementation led to a 15\% reduction in operational costs, a 10\% increase in production efficiency, a 70\% decrease in counting-related labour costs, and a 20\% reduction in lost sales. Improved filtration and model fine-tuning addressed challenges introduced by turbid water and varying fry strains. This application demonstrates the potential of using advanced computer vision and tracking algorithms to enhance commercial aquaculture operations.}

\section{Fish school behaviour analysis}
\label{sec: behaviour}
Fish behaviour, a direct result of the living environment and growth state, includes both normal (i.e. feeding behaviour, swimming behaviour, reproduction behaviour, gathering behaviour) and abnormal behaviours (i.e. disease behaviour, hypoxia behaviour, cannibalism behaviour) \cite{fotedar2016water, marques2018structure, ashley2007fish}. Poor water quality and management in aquaculture can cause fish stress behaviour \cite{li2021recent}, leading to immune suppression, slow growth, and reduced productivity and welfare \cite{ kawamura2015fish}. Traditional fish behaviour analysis, relying on human observers, is often unreliable, time-consuming, and labour-intensive \cite{an2021survey, duarte2009measurement}. Accurate estimation of fish behaviour is crucial for optimizing resource use, controlling water quality, and improving fish welfare and economic benefits \cite{liu2022nonintrusive}. The following subsections will explore the latest advancements in fish behaviour analysis, providing insights into the current state of the art and potential future directions for research and application in this field.
%focusing on computer vision-based methods for assessing fish school behaviour and feeding behaviour,

\begin{figure*}[b]
    \centering
    \includegraphics[scale=0.3]{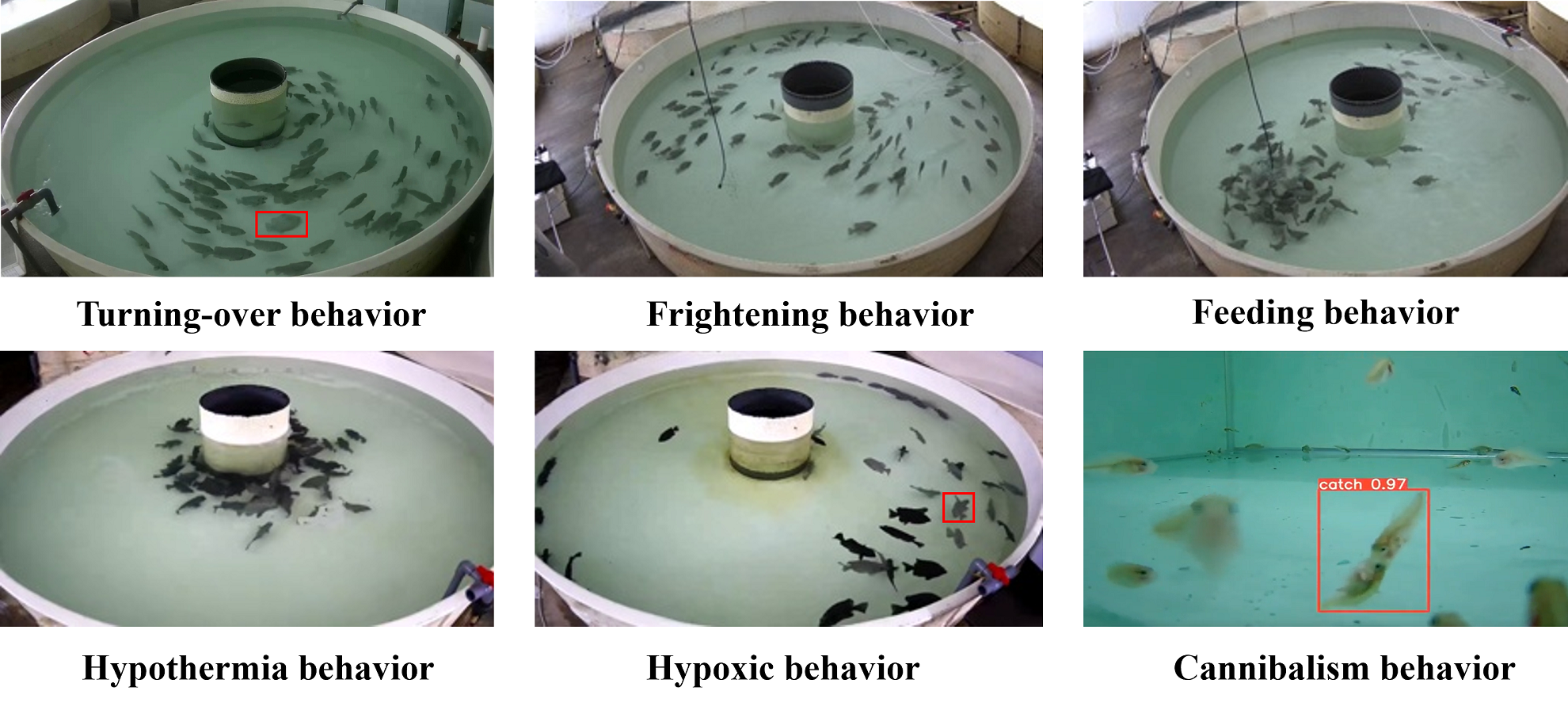}
    \caption{Abnormal behaviors: ``\textit{Turning-over behavior}'', ``\textit{Frightening behavior}'', ``\textit{Feeding behavior}'', ``\textit{Hypothermia behavior}'', ``\textit{Hypoxic behavior}'',
    ``\textit{Cannibalism behavior}''.}
    \label{fig:abnormal_behavior}
    %\vspace{-1.3em}
\end{figure*}

\subsection{Fish school behaviour analysis based on computer vision}

\subsubsection{Fish feeding behavior}

In intensive aquaculture, feeding is the main expenditure \cite{hu2021real}, and feeding optimization is crucial for improving efficiency and reducing costs \cite{sun2016models}. Traditional feeding methods based on farmers' experience are limited by low efficiency and high labour intensity, and they cannot accurately address the problems of overfeeding or underfeeding \cite{li2020automatic}. The intensity and amplitude of changes in fish behaviour can directly reflect fish appetite. Computer vision technology can effectively quantify fish feeding behaviour, optimize feeding strategies, and reduce feeding costs.

Many researchers have used traditional methods, such as background subtraction and optical flow, to extract target features for determining feeding indices \cite{zhou2017near}. While these methods can accurately capture fish feeding behaviour, they require complex foreground segmentation processes that may decrease computational efficiency and are easily affected by water surface fluctuations and reflective areas \cite{zhou2018near}. With its advantages of automatic feature extraction and large-capacity modelling, deep learning has been widely used in aquaculture \cite{feng2022fish}.

Existing approaches mainly use digital cameras to capture the corresponding images as input and characterize the fish behaviour with discrete feeding intensity (e.g., ``\textit{None}'', ``\textit{Weak}'', ``\textit{Medium}'' and ``\textit{Strong}'' \cite{ubina2021evaluating, zhou2019evaluation, zhang2023msif, yang2021dual}) as a classification problem, modelled by CNNs. However, fish-feeding behaviour is a dynamic and continuous process. Single images are insufficient to capture the context of fish feeding intensity \cite{feng2022fish, maaloy2019spatio}. As an alternative, video-based methods have been proposed to exploit spatial and temporal visual information for fish feeding intensity assessment (FFIA), which offers rich context for capturing fish feeding behaviour. Raw RGB videos were converted into optical flow image sequences and fed into a 3D CNN to evaluate fish feeding intensity, achieving a very high accuracy \cite{su2020visual, wei2021behavioral}.

While recent advancements in computer vision and deep learning have shown promise in analyzing fish feeding behaviour, some limitations still need to be addressed. One major challenge is the discrepancy between the ideal environments in which fish-feeding datasets are collected and the real-world conditions found in aquaculture settings. Factors such as water turbidity, fluctuating light levels, and variable camera angles can significantly impact the performance of these models when deployed in real-world farms.

Another limitation is the computational complexity of video-based models, which often require substantial computational resources, making them difficult to deploy on resource-constrained devices commonly used in aquaculture. The large size of these models can also hinder their real-time performance, which is crucial for timely decision-making in aquaculture management.
Furthermore, the limited generalizability of current models to new fish species is a significant challenge. Many existing models are trained on species-specific datasets, and their performance often drops significantly when applied to new or unseen species due to differences in morphological features, colour patterns, and behavioural characteristics.

To address these limitations, future research should focus on developing more robust, adaptable, and species-agnostic models that can effectively handle the variability encountered in real aquaculture environments. This may involve collecting more diverse and representative datasets, exploring domain adaptation, transfer learning, and few-shot learning techniques, and optimizing models for efficient inference on edge devices.

\subsubsection{Hypoxia behavior}

Hypoxia, a common issue in aquaculture systems, can significantly impact fish mortality and lead to substantial production losses \cite{mcfarlane2004can}. Fish exhibit various behavioural responses to hypoxic conditions, such as changes in ventilatory frequency (VF), swimming activity, surface respiration, and vertical habitat \cite{stierhoff2003hypoxia, taylor2001physiological, israeli1996monitoring, nilsson1993anoxic}. To provide early warning of hypoxia in aquaculture, it is essential to evaluate the specific behavioural responses of fish when oxygen levels in the water drop sharply.

Image processing algorithms have been proposed to quantify the hypoxia behaviour of fish in aquariums \cite{wang2021automatic}. However, these methods often rely on complex foreground segmentation processes, which can decrease computational efficiency and are easily affected by water surface fluctuations. Deep learning methods, such as YOLO object detection, have emerged as powerful tools for transforming and upgrading fish farming practices by quickly detecting fish behaviour with high accuracy \cite{hu2021real}.

Despite the progress made in recognizing fish hypoxia behaviour, most experiments have been conducted under laboratory conditions, which may not accurately reflect the challenges encountered in actual production systems. Factors such as water turbidity, uneven illumination, and high fish density can make it more difficult to identify individual fish and their specific behaviours in real-world settings. Furthermore, inducing hypoxia through human intervention in laboratory experiments can compromise animal welfare and cause irreversible damage to fish health.

To address these limitations, future research should focus on developing more robust and adaptable methods for detecting fish hypoxia behaviour in real-world aquaculture systems. Moreover, integrating multiple data sources, such as water quality sensors and video monitoring systems, could provide a more comprehensive understanding of fish behaviour and enable early detection of hypoxia-related issues. By combining advanced computer vision techniques with domain expertise in aquaculture and fish physiology, researchers can develop more effective and practical solutions for monitoring and managing fish health in real-world settings.

\subsubsection{Other abnormal behavior}

Abnormal fish behaviours, such as aggression, fear, stress, illness, parasitic infection, and cannibalism, can have significant impacts on aquaculture production (as shown in Fig. \ref{fig:abnormal_behavior}), fish welfare, and population balance \cite{ashley2007fish, frye2021cannibalism, riesch2022resource, andersson2021linking}. While less common than feeding and hypoxia behaviours, these abnormalities still play a crucial role in aquaculture warning operations. Detecting and localizing abnormal behaviours, particularly those occurring within small groups or individuals, remains challenging in computer vision.
To address this challenge, researchers have adapted techniques from human behaviour analysis, such as motion-effect maps and deep learning algorithms, to detect, localize, and recognize abnormal fish behaviours in intensive aquaculture systems \cite{zhao2018modified, wang2022fast}. These methods have shown promising results in identifying specific behaviours and evaluating various health and environmental factors. However, further research is needed to investigate the complex interplay between local and global abnormal behaviours and develop robust, multi-target tracking systems that operate efficiently in real-world aquaculture settings.

Monitoring and protecting fish during critical life events, such as spawning aggregations, is essential for maintaining population balance and preventing overfishing \cite{erisman2017fish, sadovy2005aggregation}. Computer vision techniques, including stereoscopic video analysis and 3D neural networks, have been employed to quantify fish reproductive behaviour and classify complex behaviours \cite{rastoin2020diver, long2020automatic}, providing valuable tools for baseline studies and long-term monitoring.

While computer vision and image processing technologies offer economical and effective means for monitoring abnormal fish behaviour, the relative scarcity of abnormal behaviour data has hindered in-depth research. Most existing studies have been conducted in controlled laboratory environments, which may not accurately represent the complex factors in real-world aquaculture settings \cite{piedrahita2003reducing, verdegem2006reducing}. Overcoming the challenges posed by complex water environments, uneven lighting, large numbers of individuals, and intricate fish movements are crucial for developing robust and reliable abnormal behaviour monitoring and tracking systems \cite{wang2022real}.

\subsection{Fish behaviour analysis based on multi-object tracking}

\textcolor{black}{Visual-based monitoring systems for detecting abnormal fish behaviour often rely on known scenes and predefined movement models, which can be subjective and lack adaptability to different environments \cite{francisco2020high}. Advanced tracking methods, however, offer more robust and adaptable solutions for analysing fish behaviour in diverse aquaculture settings \cite{maaloy2019spatio}.}

\textcolor{black}{Tracking technologies enable researchers to obtain comprehensive data on fish movement, including average, maximum, and minimum speeds, acceleration, average collision frequency, and trajectory changes. For example, studies have tracked zebrafish using YOLOV2 \cite{wang2023comprehensive} and Kalman filters, obtaining movement trajectories that showed significantly faster swimming, greater agitation, and agglomeration in the centre of the aquarium during feeding periods \cite{barreiros2021zebrafish}. Similarly, semi-automatic in situ tracking systems have been developed to reconstruct synchronized 3D movement trajectories of individual reef fish in social groups, analysing their behaviour when capturing plankton prey \cite{engel2021situ}.}

\textcolor{black}{Recent advancements in tracking algorithms have expanded the capabilities of fish behaviour analysis. 
Huang et al. \cite{huang2024early} developed an early warning system for detecting nocardiosis in largemouth bass (Micropterus salmoides) based on YOLOv8 \cite{sohan2024review} and ByteTrack \cite{zhang2022bytetrack}. The system quantifies fish velocity and turning angles through trajectory analysis, enabling effective early disease prevention. Similarly, Zhao et al. \cite{haixiang2024application} employed YOLOv8+ByteTrack multi-target tracking to monitor zebrafish behaviour under various environmental stressors, extracting key movement features (e.g. speeds, average collision frequency, and trajectory changes) to identify exposure to different pollutants rapidly. Xiao et al. \cite{xiao2024yolo} enhanced the tracking performance of largemouth bass (Micropterus salmoides) by combining an improved YOLOv8 \cite{sohan2024review} with ByteTrack, incorporating trajectory confidence information. This approach improved tracking accuracy and enabled more detailed analysis of fish swimming patterns. By examining the relationship between swimming speed, swimming time, and spatial coordinates, the researchers were able to quantify swimming ability in terms of both speed and endurance.} 

\textcolor{black}{Despite the advancements in tracking algorithms, their application to abnormal fish behaviour analysis in aquaculture remains limited. This scarcity can be attributed to several factors: limited availability of open datasets on abnormal fish behaviour, the rare occurrence of such behaviours making data acquisition challenging, the complexity of fish trajectories containing multi-dimensional information (position, speed, direction), and difficulty in defining `abnormal' trajectories which may encompass multiple aspects\cite{shreesha2023fish, liu2023research}.}

\textcolor{black}{To address these challenges and advance aquaculture, future research should focus on developing specialized datasets for abnormal fish behaviour in aquaculture settings, adapting advanced tracking algorithms to the unique characteristics of fish and aquatic environments, integrating domain knowledge from aquaculture experts to better define and detect abnormal behaviours, and drawing inspiration from successful applications of anomaly detection in other fields, such as crowd and vehicle monitoring. By focusing on these areas, researchers can create more powerful and flexible models for identifying and comprehending abnormal fish behaviour in various aquaculture contexts, leveraging the full potential of advanced tracking technologies.}

\subsection{Fish behaviour analysis based on passive acoustic monitoring}

\begin{figure*}
    \centering
    \includegraphics[scale=0.4]{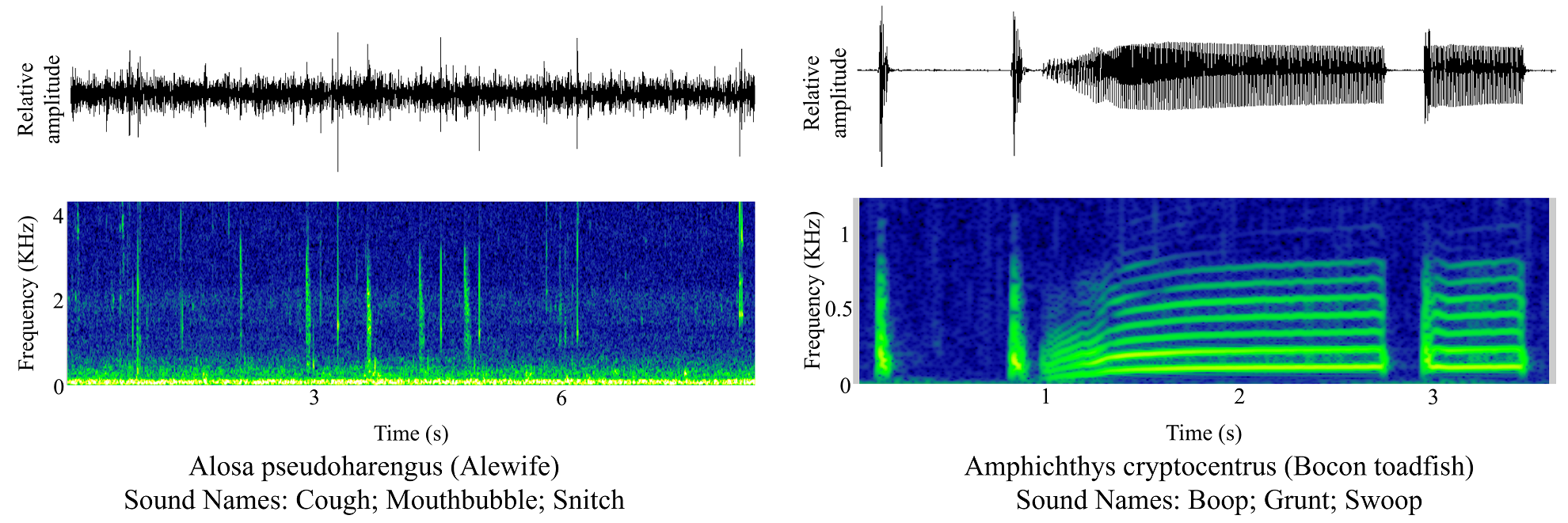}
    \captionsetup{justification=centering}
    \caption{The audio spectrum of different fish abnormal behaviours (This figure was reproduced from \cite{undersound}).}
    \label{fig:fish_spec}
\end{figure*}

Passive acoustic monitoring (PAM) has emerged as a non-invasive and increasingly accessible remote sensing technology for monitoring underwater environments \cite{chapuis2021low, parsons2022sounding}. With approximately 1,000 out of the 35,000 known fish species confirmed to produce sounds underwater \cite{froese2009fishbase, rice2022evolutionary}, PAM offers a unique opportunity to analyze fish behaviour through the sounds they generate. An audio example of fish abnormal behaviour is shown in Fig. \ref{fig:fish_spec}.

Fish can produce a series of sounds during feeding, and the frequency spectrum of these sounds can be used to analyse their feeding behaviour. For example, turbots generate feeding sounds that vary with food intake intensity, ranging from \num{15} to \num{20} dB in the frequency range \num{7}–\num{10} kHz \cite{lagardere2000feeding}. Similarly, feeding sounds produced by various fish species, such as rainbow trout (\num{0.02}–\num{25} kHz) \cite{phillips1989feeding}, Japanese minnow (\num{1}–\num{10} kHz) \cite{yamaguchi1975spectrum}, Atlantic horse mackerel (\num{1.6}–\num{4} kHz)\cite{shishkova1958notes}, yellowtail (\num{4}–\num{6} kHz) \cite{takemura1988attraction}, have comparable frequency ranges.

The Fish feeding behaviour analysis based on audio was initially proposed by \cite{cui2022fish, cui2023multimodal}, where the audio signal is first transformed into log Mel spectrograms and then fed into a CNN-based model for FFIA. Subsequent work \cite{du2023feeding, zeng2023fish} have further demonstrated the feasibility of using audio as input for FFIA. Audio-based methods offer advantages such as energy efficiency and lower computational costs compared to vision-based methods \cite{gao2018learning, gao2020listen}. However, audio-based models have lower classification performance than video-based FFIA due to their inability to capture full visual information and sensitivity to environmental noise \cite{choi2022temporal}. Moreover, rapidly swimming predatory fish, such as brown and rainbow trout, often combine forward swimming with feeding, accompanied by splashing sounds and strong tail patting \cite{liu2022nonintrusive}. The rapid pellet capture by these species superimposes feeding sounds, and pellet impacts pose a challenge in obtaining accurate feeding sound data.

To overcome these challenges, future research should focus on developing advanced signal processing techniques to separate feeding sounds from ambient noise and other interfering sounds. Additionally, exploring the integration of audio and visual data could help improve the overall classification performance and robustness of fish behaviour analysis systems. 

\subsection{Fish behaviour analysis based on biosensor technology}

Biosensor technology has shown great potential in collecting individual animal information, such as individual trajectory, acceleration, velocity, respiration frequencies, heartbeat frequency, and tail beat frequency \cite{martos2019ultra, rosell2021use}. In recent years, accelerometers have been increasingly used in marine biology research to study the feeding behaviour of aquatic animals.

The feeding behaviour of most fish leads to characteristic changes in acceleration that differ from their normal movement patterns \cite{horie2016development}. These characteristic changes in acceleration can be effectively used to distinguish feeding behaviour patterns from other behaviour patterns \cite{tanoue2012feeding}. For example, in \cite{broell2013accelerometer}, accelerometer tags were used to investigate the feeding behaviour of Atlantic cod (Gadus morhua) in the wild. The authors found that the accelerometer data could accurately identify feeding events and provide insights into the foraging ecology of this species. Similarly, a study by \cite{kawabata2014use} used a combination of accelerometers and gyroscopes to analyse the feeding behaviour of captive yellowtail kingfish (Seriola lalandi). The authors demonstrated that the sensor data could be used to classify different types of feeding behaviour, such as biting, chewing, and swallowing, with high accuracy.

In addition to feeding behaviour, biosensors have been used to study other aspects of fish behaviour, such as swimming activity and energy expenditure. For instance, in \cite{birnie2019cortisol}, accelerometers were used to investigate the swimming behaviour and energy expenditure of wild Atlantic salmon (Salmo salar) while migrating to spawning grounds. The authors found that the accelerometer data provided valuable insights into the swimming performance and energy costs of this species in natural conditions.

However, using biosensors in fish behaviour analysis also presents some challenges and concerns. Biosensors are typically surgically attached or implanted into the fish's body, which can lead to the direct death of the fish or cause behavioural changes that may affect the results of experiments. Moreover, this method may cause irreversible harm to the fish and compromise animal welfare. To address these issues, researchers should focus on developing minimally invasive or non-invasive biosensor technologies that can be safely attached to or removed from fish without causing undue stress or harm. Furthermore, ethical considerations should be prioritized when using biosensor technology in fish behaviour analysis. 

Despite these challenges, biosensor technology offers a promising approach to studying fish behaviour at the individual level, providing valuable insights into the feeding ecology, swimming performance, and energy expenditure of various fish species. By combining biosensor data with other monitoring techniques, such as passive acoustic monitoring and vision-based methods, researchers can develop a more comprehensive understanding of fish behaviour in both captive and wild settings. As biosensor technology continues to advance, it is essential to balance the potential benefits of these tools with the need to ensure the welfare and ethical treatment of the fish being studied.

\subsection{Comparative analysis and trends in fish school behaviour analysis}
\textcolor{black}{Fish school behaviour analysis in aquaculture focuses on understanding and monitoring fish activities, stress responses, and interactions within the aquaculture environment. This field employs various technologies, each with distinct capabilities and applications.
Computer vision-based analysis excels in observing feeding patterns, detecting abnormal behaviours, and identifying signs of stress or disease \cite{an2021application}. Advanced algorithms can recognize specific behaviour patterns such as aggressive interactions, schooling dynamics, and responses to environmental changes. While being non-invasive, this method's effectiveness can be limited by water clarity and lighting conditions \cite{shreesha2020computer}.
Multi-object tracking extends computer vision capabilities by following individual fish within schools over time. This approach provides insights into social hierarchies, individual growth rates, and how behaviour patterns spread through a population. It requires sophisticated algorithms and high computational power but offers unparalleled detail in understanding group dynamics \cite{delcourt2013video, an2021survey}.}

\textcolor{black}{
Passive acoustic monitoring analyses sounds produced by fish to infer behaviour states and stress levels \cite{chapuis2021low}. This method is particularly valuable for nocturnal species or in turbid environments where visual observation is challenging. It can detect feeding activity, spawning events, and even some types of distress calls, offering a unique perspective on fish behaviour \cite{luczkovich2008passive}.
Biosensor technology, including implantable tags and external sensors, provides detailed physiological data that can be correlated with behaviour \cite{broell2013accelerometer}. These sensors can measure parameters like heart rate, muscle activity, and hormone levels, offering insights into stress responses, energy expenditure during different activities, and even feeding intensity \cite{behera2023recent}. However, the invasive nature of some biosensors raises welfare concerns and may itself affect behaviour.
The choice of behaviour analysis method often depends on research goals, species-specific characteristics, and environmental conditions. For example, salmon farms might prioritize feeding behaviour analysis to optimize feed conversion ratios, while tilapia farms might focus more on aggression and territoriality behaviours.}

\textcolor{black}{
Regional trends in adoption vary. Asia, with its diverse aquaculture species and systems, shows broad adoption across methods \cite{pangsorn2022issues,eguiraun2023entropy}. Europe emphasizes non-invasive techniques due to strict welfare regulations \cite{roth2024experimental}. North America leads in integrating multiple methods for comprehensive behaviour analysis 
Welfare considerations are increasingly central to method selection \cite{kuiper2023advances, aguzzi2020potential}. There is a growing preference for non-invasive techniques that can provide detailed behavioural data without causing stress. This has spurred research into improving the resolution and accuracy of passive monitoring methods.
The trend in fish behaviour analysis is moving towards holistic approaches that combine multiple methods to create a comprehensive picture of fish welfare and behaviour. For instance, combining computer vision with passive acoustics can provide round-the-clock monitoring that captures both visual and auditory behavioural cues \cite{cui2023multimodal}.
Real-world applications have demonstrated the value of behaviour analysis in improving aquaculture management. Farms using advanced behaviour monitoring have reported earlier detection of disease outbreaks, more efficient feeding practices, and improved overall fish welfare, leading to better growth rates and product quality \cite{fore2018precision, wang2021intelligent}.
As aquaculture intensifies and faces new challenges like climate change impacts, behaviour analysis will play an increasingly crucial role in maintaining fish health, optimizing production, and ensuring sustainable practices.}

\subsection{Real-world applications in fish behaviour analysis}
\textcolor{black}{Fish behaviour analysis in real-world applications has emerged as a powerful tool for monitoring aquaculture environmental conditions and fish welfare. By tracking and analysing changes in swimming patterns, spatial distribution, and movement characteristics, these systems can serve as early warning indicators for water quality issues and stress responses, enabling timely interventions in commercial aquaculture operations. Xu et al. \cite{xu2024behavioral} demonstrate how fish behaviour analysis can be used to monitor water quality by studying responses to ammonia nitrogen stress. Their system combines an improved YOLOv8 model (enhanced with multi-head self-attention and Wise Intersection over Union (Wise-IoU)) with dual-view cameras for 3D tracking. By integrating advanced tracking techniques including Kalman filtering and Kernelized Correlation Filters \cite{henriques2014high}, the system accurately monitors multiple fish simultaneously. The study analysed behavioural patterns of different species (sturgeon, bass, and crucian) under various ammonia nitrogen stress levels, examining their trajectories, exercise volumes, spatial distribution, and movement velocities.}

% Xu et al. \cite{xu2024behavioral} introduces an innovative approach to monitoring water quality in aquaculture by analyzing fish behaviour under ammonia nitrogen stress using deep learning and 3D movement trajectory analysis. The researchers developed an improved YOLOv8 model enhanced with multi-head self-attention and Wise-IoU for object detection, addressing challenges such as occlusion and high-density scenarios in fish tracking. To overcome the limitations of single-camera setups and refraction issues in underwater environments, they implemented a dual-view camera system with careful calibration. The method combines the enhanced YOLOv8 model with 3D positioning techniques using the Kalman filter, Kuhn-Munkres algorithm, and Kernelized Correlation Filters. This integration allows for accurate tracking of multiple fish simultaneously, even in complex aquarium environments. To handle the small size and fast movement of juvenile bass, which initially posed detection difficulties, the researchers applied data augmentation techniques and optimized the channel design to encourage fish separation. The study analyzed the behaviour of sturgeon, bass, and crucian under various ammonia nitrogen stress levels, evaluating their trajectories, volumes of exercise, spatial distribution, and movement velocity.

\begin{figure}[t]
    \centering
    \includegraphics[width=\linewidth]{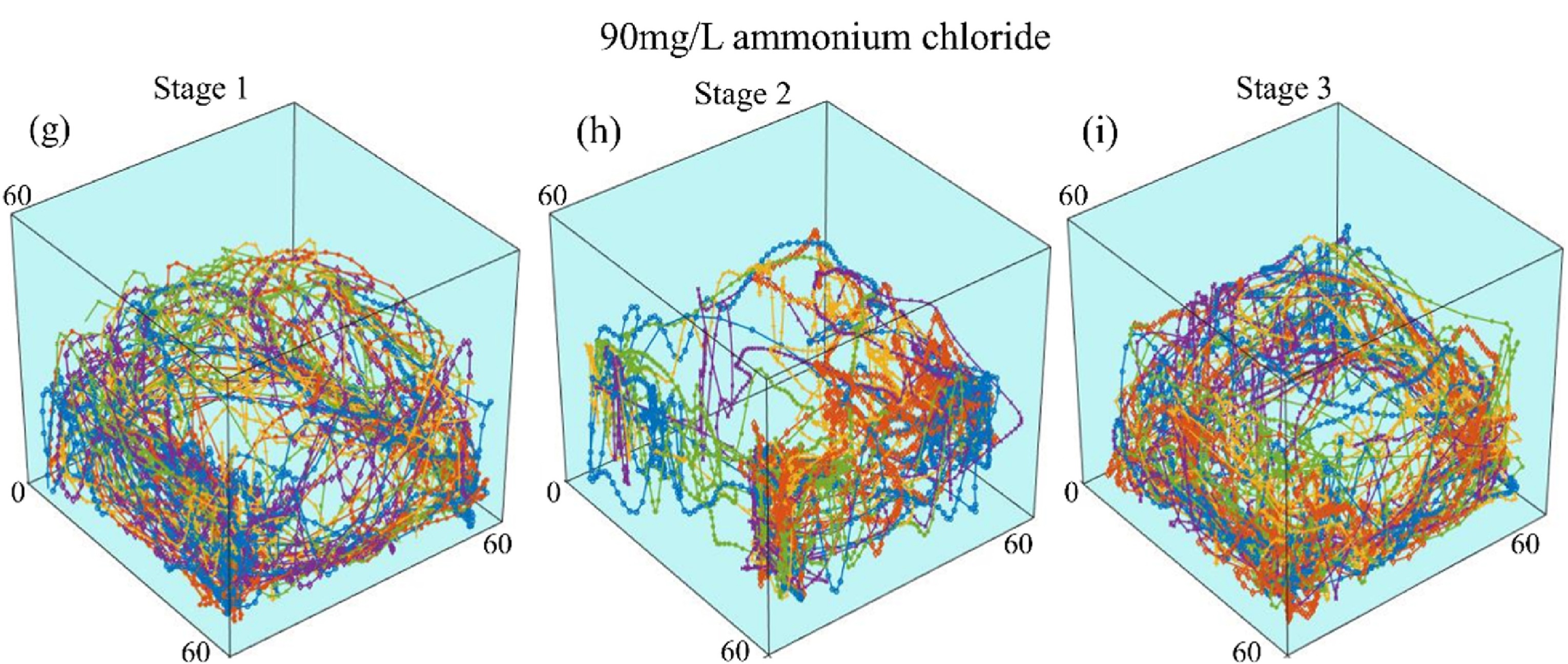}
    \caption{Behavioral trajectories of sturgeon under ammonia nitrogen stress. The five colours of the behavioural trajectory represent five fish (This figure was reproduced from \cite{xu2024behavioral}).}
    \label{fig: abnormal}
\end{figure}

\textcolor{black}{Fig. \ref{fig: abnormal} shows that sturgeons swim around the aquarium under normal conditions while they swim up and down at a certain position in the aquarium under ammonia nitrogen stress \cite{xu2024behavioral}. Meanwhile, the phenomenon of sturgeon swimming around the aquarium under ammonia nitrogen stress has decreased. Key findings revealed significant changes in fish swimming patterns, spatial distribution, and velocity under ammonia stress, with species-specific responses to different concentrations. The method successfully differentiated normal behaviour from stress-induced abnormalities, demonstrating its potential for early detection of water quality issues in aquaculture settings. However, challenges such as turbid water and varying light conditions required additional solutions like adaptive image preprocessing techniques and regular maintenance protocols.}
% However, challenges such as turbid water and varying light conditions required additional solutions. The researchers implemented adaptive image preprocessing techniques and scheduled regular camera cleaning to maintain detection accuracy in less-than-ideal conditions. These adaptations highlight the method's robustness and potential for real-world applications in diverse aquaculture environments.

\section{Multimodal fusion in aquaculture}
\label{sec:multimodal}
\textcolor{black}{Multimodal data fusion has been used in aquaculture where data from various sources are integrated to understand fish behaviour, health, and environmental conditions. Recent advancements in other fields, such as deep learning, offer promising applications for aquaculture, particularly in fish tracking, counting, and behaviour analysis.} 

\subsubsection{Multimodal fusion in fish tracking}
\textcolor{black}{For fish tracking, state-of-the-art vision-language models show significant potential. CLIP2Video \cite{fang2021clip2video} and AudioCLIP \cite{guzhov2022audioclip} extend the capabilities of their predecessors in image-text and audio-text matching by leveraging the contrastive language image pretraining (CLIP) model \cite{radford2021learning}, and could be adapted for multimodal fish detection. For instance, the AudioCLIP architecture \cite{guzhov2022audioclip} could be modified to correlate visual fish features with sonar data, potentially improving detection accuracy in turbid waters commonly encountered in aquaculture settings. However, the significant differences between underwater RGB images and sonar images present unique challenges in multimodal tracking for aquaculture. Many existing multimodal tracking methods assume alignment between different modal spaces, using a prediction head to estimate the same target box across modalities. This assumption often fails in RGB-Sonar (RGB-S) tracking, where the representation of the target can vary significantly between the two imaging types \cite{lu2023image, qadri2024aoneus}. Furthermore, the spatial misalignment of multimodal features complicates the integration of information from these diverse sources. A critical challenge in multimodal fusion for aquaculture is data synchronization. Different sensors often have varying sampling rates and latencies. For instance, acoustic sensors might collect data at several kilohertz (kHz), while visual sensors typically capture frames at 30-60 Hz, and environmental sensors (like temperature or pH sensors) might only take measurements every few seconds or minutes.}
% acoustic sensors might provide data at a sampling rate often different from visual sensors, while environmental sensors may update less frequently. 

\textcolor{black}{Developing robust synchronization methods, such as timestamp alignment and interpolation techniques, is crucial for accurately fusing these diverse data streams in real-time aquaculture monitoring systems.
To address these challenges, researchers have developed specialized approaches for underwater tracking. Li et al. \cite{li2024rgb} introduced a novel spatial cross-attention method that effectively bridges the gap between underwater RGB and sonar modalities. Their approach, which builds upon the Cross-Attention method \cite{vaswani2017attention}, enables more nuanced interactive fusion between the different modalities. By accounting for the unique characteristics of each imaging type, this method achieves state-of-the-art performance in underwater tracking, demonstrating its potential for improving fish monitoring in complex aquaculture environments. The introduction of large-scale datasets like WebUOT-1M \cite{zhang2024webuot}, which includes language prompts for video sequences, opens up new possibilities for underwater vision-language tracking. This could be particularly useful for developing more robust fish tracking systems in complex aquaculture environments, such as sea cages with varying light conditions and water clarity. However, a significant research gap remains in bridging vision-language models with sonar data for aquaculture applications, particularly in developing frameworks that can effectively leverage textual descriptions for enhanced underwater object detection and tracking. Future research should focus on developing more sophisticated multimodal fusion techniques to improve tracking accuracy and robustness in complex underwater environments.}

\subsubsection{Multimodal fusion in fish counting}
 \textcolor{black}{In the realm of fish counting, recent developments in crowd counting using vision-language models offer promising approaches. CrowdCLIP \cite{liang2023crowdclip} leverages pre-trained vision-language models to address the challenge of costly manual labelling, especially in dense scenes. This approach could be adapted for counting fish in crowded aquaculture tanks or ponds, where traditional methods often fail. CLIP-Count \cite{jiang2023clip} presents an innovative end-to-end pipeline for estimating density maps of open-vocabulary objects with text guidance in a zero-shot manner. Its patch-text contrastive loss and hierarchical patch-text interaction module could be adapted to generate high-quality density maps for fish in various aquaculture scenarios, potentially improving counting accuracy in complex underwater environments with overlapping fish. Furthermore, VLCounter \cite{kang2024vlcounter} explores the implicit association of semantic-patch embeddings of CLIP, showing improved performance in zero-shot object counting. This approach could be useful in aquaculture for counting different fish species or identifying specific fish features without extensive labelled datasets, a common challenge in diverse aquaculture operations. However, adapting these vision-language models to underwater environments presents unique challenges, particularly in handling water turbidity and varying lighting conditions. Future research should focus on developing specialized vision-language architectures that can better handle the specific challenges of underwater counting tasks while maintaining the benefits of zero-shot learning capabilities.}
 
\subsubsection{Multimodal fusion in fish behaviour analysis}
\textcolor{black}{In fish behaviour analysis, multi-sensor approaches combined with advanced algorithms offer new insights. The Multimodal Foundation Model (M3AE) \cite{geng2022m3ae}, designed for general-purpose multimodal understanding, could be adapted to integrate visual, acoustic, and environmental data for comprehensive fish behaviour monitoring. This integration could enable early detection of stress or disease by correlating subtle behavioural changes with environmental parameters. Cui et al. \cite{cui2023multimodal} introduced a novel unified mixed-modality-based method for fish feeding behaviour assessment, capable of processing audio, visual, or audio-visual modalities. This approach demonstrates the potential of multimodal fusion to provide more robust and accurate assessments of fish behaviour than single-modality based methods. Similarly, Du et al. \cite{du2024harnessing} showed that fusing information from audio, video, and imaging sonar modalities resulted in higher accuracy than using a single modality in assessing fish behaviour. While these studies demonstrate promising results, there remains a need to develop more sophisticated behavioural pattern recognition algorithms that can identify complex fish behaviours across different species and environmental conditions. Future research should explore integrating biological knowledge with multimodal deep learning approaches to better understand and interpret fish behavioural patterns in various aquaculture scenarios.}

\subsubsection{Challenges and solutions in multimodal fusion for aquaculture}
\textcolor{black}{Noise management is particularly challenging in underwater environments. Each modality (visual, acoustic) is subject to different types of noise. For example, visual data may be affected by turbidity and light scattering, while acoustic data can be distorted by ambient noise from equipment or other marine life. Advanced filtering techniques, such as Bernoulli filters \cite{zhao2022audio} and differentiable particle filtering \cite{qian2019multi}, need to be developed specifically for aquaculture applications to ensure reliable data fusion in these complex environments \cite{yang2024cmaf}. The relative importance and reliability of different data sources can vary based on environmental conditions and the specific behaviour being analyzed. Developing dynamic weighting schemes that adjust the contribution of each modality based on real-time quality assessments is essential \cite{chen2024uamfdet}. For instance, acoustic data might be weighted more heavily in turbid conditions than visual data, while in clear water, the reverse might be true. Machine learning techniques, such as attention mechanisms in transformer-based models, offer promising approaches for adaptive weighting in multimodal fusion systems.}

\textcolor{black}{However, several challenges limit the direct application of these methods in aquaculture. Limited labelled data is a significant hurdle, as aquaculture-specific datasets are scarce compared to general computer vision tasks. Environmental variability in underwater conditions, such as changing turbidity, lighting, and temperature, can significantly impact sensor reliability and model performance. Additionally, species-specific behaviours require specialized models, as different fish species exhibit unique patterns  \cite{li2024review}. The dynamic nature of fish movement and the three-dimensional space of aquatic environments further complicate the application of these technologies. Several approaches have shown promise in addressing these challenges. Few-shot learning techniques could help the models adapt to new species or environments with limited data \cite{jin2017deep}, crucial for the diverse range of fish species in aquaculture. Self-supervised learning methods \cite{saleh2024track,tarling2022deep} could leverage large amounts of unlabeled aquaculture data to pre-train robust models, potentially overcoming the scarcity of labelled datasets. Domain adaptation techniques \cite{zhao2022unsupervised} could help transfer knowledge from data-rich domains to aquaculture-specific tasks, bridging the gap between general computer vision tasks and the unique requirements of underwater environments.}

\textcolor{black}{Future research should focus on developing specialized datasets that capture the diversity of aquaculture environments and species. This could include multi-sensor data collected across different water conditions, fish life stages, and farming systems. Adapting advanced algorithms to underwater environments is crucial, particularly in handling variable lighting, water turbidity, and three-dimensional fish movement. Creating robust, generalizable models for diverse aquaculture scenarios will be key to the widespread adoption of these technologies. Integrating multiple data sources and advanced processing techniques can revolutionize fish monitoring, enhance production efficiency, and improve animal welfare in aquaculture operations. However, it is important to note that applying these technologies in real-world aquaculture settings may face practical challenges such as cost, maintenance of underwater sensors, and the need for real-time processing capabilities. As research progresses, critically evaluating the practicality and cost-effectiveness of these advanced multimodal approaches in various aquaculture settings will be essential.}
\section{Public dataset}
\label{sec: dataset}
\textcolor{black}{High-quality public datasets are crucial for developing and evaluating various methods including deep learning methods for fish detection, tracking, and behaviour analysis. Despite growing efforts, public datasets on underwater fish scenes are scarce, particularly in realistic aquaculture environments. This limitation has led many researchers to conduct analyses and behavioural studies under ideal or controlled conditions, which may not fully represent the challenges of real-world aquaculture settings. Table \ref{tab: dataset} summarizes the available public fish datasets, highlighting their diverse characteristics and limitations.}

\begin{table*}[]
\caption{ Summary of the various fish datasets.}
\label{tab: dataset}
\centering
\begin{tabular}{@{}cccccc@{}}
\toprule
Dataset                                                                                                       & \begin{tabular}[c]{@{}c@{}}No. of \\ videos/image\end{tabular}                                                         & Resolution                                                                          & \begin{tabular}[c]{@{}c@{}}Number of \\ labeled data\end{tabular} & Tasks                                                                                   & Reference                                                                                                   \\ \midrule
\begin{tabular}[c]{@{}c@{}}Fish4- \\ Knowledge\end{tabular}                                                   & \begin{tabular}[c]{@{}c@{}}700,000 videos \\ with $10 \mathrm{~min}$ \\ each clip\end{tabular}                         & $320 \times 240$                                                                    & -                                                                 & \begin{tabular}[c]{@{}c@{}}Classification, \\ Detection and \\ Tracking\end{tabular}    & 
\cite{Fish4-Knowledge}                                                                    \\
SeaCLEF 2016                                                                                                  & \begin{tabular}[c]{@{}c@{}}Training set: 20 \\ videos and \\ 20,000 images, \\ Test set: 73 \\ videos\end{tabular}     & \begin{tabular}[c]{@{}c@{}}$640 \times 480$ \\ $320 \times 240$\end{tabular}        & 9,000                                                              & \begin{tabular}[c]{@{}c@{}}Classification, \\ Counting\end{tabular}                     &\cite{SeaCLEF20}                                                           \\
NCFM                                                                                                          & \begin{tabular}[c]{@{}c@{}}16,915 images \\ $(3,777$ training, \\ 13,138 testing)\end{tabular}                         & $1920 \times 1080$                                                                  & 10000                                                             & \begin{tabular}[c]{@{}c@{}}Detection, \\ classification and \\ counting\end{tabular}    & \begin{tabular}[c]{@{}c@{}} \cite{NCFM}  \end{tabular}  \\

\begin{tabular}[c]{@{}c@{}}Sonar image \\ counting dataset\end{tabular}                                       & \begin{tabular}[c]{@{}c@{}}30 videos \\ sequence with \\ 537 images\end{tabular}                                       & $360 \times 360$                                                                    & -     & Counting                                                                                & \cite{Sonacounting}                                                                         \\
3D-ZeF20                                                                               & \begin{tabular}[c]{@{}c@{}}Training Set: \\ 54052 images, \\ Test Set: 32400 \\ images\end{tabular}                    & $2704 \times 1520$                                                                  & 86,452                                                             & Tracking                                                                                & \cite{pedersen20203d}                                                                    \\
FishTracker23                                                                               & \begin{tabular}[c]{@{}c@{}}Approaching 1 million\\ boxes and frames\end{tabular}                    &  \begin{tabular}[c]{@{}c@{}}$1920 \times 1080$   \\$1280 \times 720$  \end{tabular}                                                              &  \begin{tabular}[c]{@{}c@{}}850,000 \end{tabular}                                                             & Tracking                                                                                & \cite{dawkins2024fishtrack23}   
                                   \\
WebUOT-1M                                                                              & \begin{tabular}[c]{@{}c@{}}1,500 underwater \\ videos (total 10.5 hours) \\and 408 categories\end{tabular}                    & -                                                                  & 1,100,000                                                             & Tracking                                                                                & \cite{zhang2024webuot} \\
\begin{tabular}[c]{@{}c@{}}Automated\\Fish Tracking \end{tabular}                                                                                                     &   \begin{tabular}[c]{@{}c@{}}189 videos of\\ varying durations\\(1- 30 seconds) \end{tabular}                                                                                                        & $1920 \times 1080$                                                                  & 8,700                                                       & \begin{tabular}[c]{@{}c@{}}Detection, \\ Tracking\end{tabular} & \cite{lopez2021automatic}  \\                                   DeepFish                                                                                                      & 39,766 images                                                                                                          & $1920 \times 1080$                                                                  & 3200                                                       & \begin{tabular}[c]{@{}c@{}}Segmentation, \\ counting and \\ Classification\end{tabular} & \cite{saleh2020realistic}                             \\
FISHTRAC                                                                                                      & 14 videos                                                                                                              & $1920 \times 1080$                                                                  & 3,449                                                              & \begin{tabular}[c]{@{}c@{}}Tracking and \\ detection\end{tabular}                       & \begin{tabular}[c]{@{}c@{}}\cite{mandel2023detection}\end{tabular} \\
BrackishMOT                                                                                                   & \begin{tabular}[c]{@{}c@{}}98 videos each \\ lasting about 1 \\ minute\end{tabular}                                    & $2704 \times 1520$                                                                  &                                                                   & Tracking                                                                                & \cite{pedersen2023brackishmot}                                                                      \\
$\mathrm{CFC}$                                                                                                & \begin{tabular}[c]{@{}c@{}}527215 \\ SONA images\end{tabular}                                                          & \begin{tabular}[c]{@{}c@{}}$288 \times 624$ to \\ $1086 \times 2125$\end{tabular} & 515,933                                                            & \begin{tabular}[c]{@{}c@{}}Detection, Tracking \\ and Counting\end{tabular}             & \cite{cfc2022eccv}    \\
\begin{tabular}[c]{@{}c@{}} Mullet Schools \\ Dataset\end{tabular}                                                                                                & \begin{tabular}[c]{@{}c@{}}over 100k \\ SONA images\end{tabular}                                                          & \begin{tabular}[c]{@{}c@{}}$320 \times 576$ \end{tabular} & 500                                                            & \begin{tabular}[c]{@{}c@{}}Detection \\ Counting\end{tabular}             & \cite{tarling2022deep}   \\
\begin{tabular}[c]{@{}c@{}} Fish Sounds \end{tabular}                                                                                                & \begin{tabular}[c]{@{}c@{}}115 different fish \\ sound clips\end{tabular}                                                          & \begin{tabular}[c]{@{}c@{}}64kbps \end{tabular} & -                                                            & \begin{tabular}[c]{@{}c@{}}Behaviour \\ analysis\end{tabular}             & \cite{Fishsound}  \\
\begin{tabular}[c]{@{}c@{}} AV-FFIA \end{tabular}                                                                                     & \begin{tabular}[c]{@{}c@{}}27000 video \\and sound clips\end{tabular}                                                          & \begin{tabular}[c]{@{}c@{}}$1086 \times 2125$\\256kbps \end{tabular} & All                                                           & \begin{tabular}[c]{@{}c@{}}Feeding Behaviour \\ analysis\end{tabular}             & \cite{cui2023multimodal}
\\ \bottomrule
\end{tabular}

\end{table*}

\textcolor{black}{The Fish4Knowledge dataset \cite{Fish4-Knowledge}, captured in Taiwan waters from 2010 to 2013, offers a variety of marine scenes but suffers from low resolution (i.e. 320 × 240 pixels), limiting its applicability to current research needs.  SeaCLEF2016 \cite{SeaCLEF20} builds on this with diverse species coverage and high-quality annotations, enhancing automated marine life identification capabilities. However, both datasets have limitations, including geographic specificity to Taiwanese waters, fixed camera positions that may miss certain behaviours, and low video quality issues due to underwater conditions. The DeepFish dataset \cite{saleh2020realistic} significantly contributes to underwater fish monitoring, containing 39,766 images from 20 locations in Australian marine environments. It offers annotations for multiple tasks including classification, counting, localization, and polygonal annotations for precise fish shape delineation. While being versatile, DeepFish initially lacks segmentation masks, limiting its use in instance segmentation tasks. Its diverse environmental coverage, from clear to turbid waters, enhances its utility for developing robust algorithms. However, the dataset's focus on still images and specific marine ecosystems may limit its direct applicability to some aquaculture scenarios or dynamic behaviour studies.}

\textcolor{black}{The WebUOT-1M dataset \cite{zhang2024webuot} represents a significant advancement, offering a large-scale solution with 1.1 million frames across 1,500 underwater videos and 408 diverse target categories. Its inclusion of language prompts for each video facilitates multi-modal research and enables the development of more generalized and multi-modality underwater object-tracking models. However, the dataset's web-sourced nature means it may include some biases in image quality, object representation, and environmental conditions. While this diversity can be beneficial for creating robust models, it may not always accurately represent real-world underwater scenarios encountered in specific research or industrial applications. For specialized research, datasets like 3DZeF20 \cite{pedersen20203d} and AV-FFIA \cite{cui2023multimodal} offer unique perspectives. 3DZeF20 provides high-precision 3D data for zebrafish in laboratory settings, invaluable for detailed behaviour analysis but limited in ecological applicability. AV-FFIA innovatively combines hydrophone and video data for fish feeding behaviour analysis, offering deep insights into specific patterns. While all of them are invaluable for specific research questions, their applicability to broad aquaculture scenarios may be limited due to their focus on controlled environments or single species.}

\textcolor{black}{FishTrack23 \cite{dawkins2024fishtrack23} and FISHTRAC \cite{mandel2023detection} represent the current state-of-the-art in real-world underwater object detection and tracking. These datasets address the complexities of natural marine environments, providing comprehensive annotations for multiple tasks. They excel in developing practical monitoring tools but may still face challenges with image quality consistency and global species representation. The BrackishMOT dataset\cite{pedersen2023brackishmot} uniquely addresses tracking in turbid environments, which is crucial for many aquaculture settings. The Caltech Fish Counting (CFC) dataset \cite{cfc2022eccv} significantly contributes to underwater computer vision and marine ecology and is specifically designed to address the challenges of fish counting in complex underwater environments. Its diverse scenes capture different lighting conditions, water turbidity levels, and fish densities, making it an excellent resource for developing robust fish detection and counting algorithms. The dataset's strength lies in its real-world applicability, presenting researchers with the genuine challenges faced in marine population surveys, such as occlusions, varying fish sizes, and species diversity. However, users should be aware that while the CFC dataset offers a comprehensive view of coral reef fish populations, it may have limitations in representing global marine ecosystems. The Fishsounds dataset \cite{Fishsound}, while offering audio data for various fish species, lacks sufficient data for comprehensive behaviour analysis. The Mullet Schools dataset \cite{tarling2022deep} provides large-scale sonar imagery for fish counting but is species-specific, limiting its broad applicability in aquaculture research.}

\textcolor{black}{These datasets highlight several limitations in current public resources for aquaculture research. Many focus on specific conditions or species, not capturing the full range of aquaculture environments and diversity. Most are unimodal, lacking the integration of visual, acoustic, and environmental data crucial for comprehensive aquaculture monitoring. Varying annotation standards across datasets complicate cross-dataset evaluations. We suggest creating multi-modal datasets that integrate visual, acoustic, and environmental data specific to aquaculture settings to address these limitations and better serve future research needs. Existing datasets should be expanded to include a wider range of species and environmental conditions relevant to global aquaculture practices. Developing standardized annotation protocols would ensure consistency across aquaculture-focused datasets. Establishing benchmark datasets specifically designed for aquaculture applications, covering various farming systems and environmental conditions, would greatly benefit the field.}
\section{Challenges and future perspectives}
\label{sec: challenges}
Fish tracking, counting, and behaviour analysis play a crucial role in the intelligent development of aquaculture production. While computer vision technology is currently a popular method for these tasks, it faces several challenges due to the unique characteristics of aquaculture environments, such as high fish density, complex water backgrounds, and irregular fish movement. These factors can lead to interference between multiple targets, false detections, missed counts, and tracking failures.

Acoustic methods offer an alternative approach that enables automatic and rapid fish counting and tracking in low-light and turbid water conditions. However, underwater noises, high equipment costs, and the need for professional expertise make acoustic methods more suitable for large-scale operations like marine fishing rather than factory or pond farming environments.
To further increase the level of intelligence in aquaculture, we predict several different trends for future development:

1) Massively available datasets: The wide application of intelligent technology in aquaculture, especially the success of deep learning algorithms in image processing \cite{yang2021deep}, has highlighted the need for large labelled datasets. Although available datasets are gradually increasing, most are limited to identifying and detecting fish species. Open data on fish tracking, counting, and behaviour analysis is scarce. Passive acoustic monitoring is also gaining popularity for underwater listening \cite{lin2021exploring, parsons2022sounding} and public sound data of underwater fish (e.g., Fishsound) are emerging. However, the sample size of these datasets has not yet reached critical mass. In the future, developing an international platform for sharing images and acoustic data will be essential to promote sustainable aquaculture development.

2) Edge computing and real-time processing:
\textcolor{black}{As aquaculture operations become more technologically advanced, there is an increasing need for real-time data processing and decision-making at the edge. Developing efficient algorithms and hardware solutions that can perform complex tasks like fish tracking, counting, and behaviour analysis on-site, without relying on cloud computing, will be crucial. This approach can reduce latency, improve data privacy, and enable faster responses to changing conditions in aquaculture environments. Future research should optimise existing algorithms for edge devices, develop specialized hardware for aquaculture applications, and create integrated systems that seamlessly combine multiple monitoring and analysis tasks in real-time.}

3) On-device machine learning: Most current fish tracking, counting, and behavioural analysis models are performed in the cloud or on high-performance GPUs. However, many aquaculture tasks require real-time responses, such as fish feeding and abnormal behaviour detection. Cloud-based models may struggle to guarantee this real-time performance, and many devices in remote and harsh aquaculture environments may not have consistent internet connectivity. On-device models can greatly reduce exercise pressure and make devices more intelligent, providing users with a better experience. However, terminal devices are often limited in processing power, power consumption, cost, and volume. Future developments could focus on reducing the complexity of computing and storage by optimizing neural network algorithms or compressing network models using techniques like knowledge distillation to enable their deployment on device chips.

4) Integration of fish tracking, counting, and behaviour analysis:
Most research addresses fish tracking, counting, and behaviour analysis as separate tasks. However, these tasks are often interconnected in real-world aquaculture scenarios and must be performed continuously in the same environment. Developing a joint model that can handle all three tasks simultaneously would be more memory-efficient and suitable for practical applications in aquaculture.
A joint model would leverage the shared features and information among the tasks, reducing redundancy and improving overall performance. For example, accurate fish tracking can provide valuable counting and behaviour analysis information. In contrast, behaviour analysis can help identify and resolve tracking challenges such as occlusions and interactions between fish.
More comprehensive and efficient systems could be developed for monitoring and managing aquaculture farms by integrating these tasks into a single framework. This approach would also reduce the computational resources required, making it more feasible to deploy such systems in real-world settings.
Future research should focus on developing novel architectures and training strategies that can effectively combine fish tracking, counting, and behaviour analysis tasks.

5) Integration of large language models (LLMs) and artificial general intelligence (AGI): Recent advancements in LLMs and AGI have the potential to revolutionize fish tracking, counting, and behaviour analysis. LLMs, such as GPT-4 \cite{GPT4} and LLaMA \cite{llama}, can be fine-tuned on aquaculture-specific datasets to generate accurate descriptions and analyses of fish behaviour from textual data. AGI systems, like DeepMind's Gato \cite{reed2022generalist}, which can perform a wide range of tasks using a single model, could be adapted to integrate multiple modalities (e.g., vision, acoustics, and text) for comprehensive fish monitoring and management. By leveraging the power of LLMs and AGI, aquaculture researchers and practitioners can develop more intelligent and adaptable systems for understanding and optimizing fish welfare and production.
\section{Conclusions}
\label{sec: conclusion}
This survey provides a comprehensive analysis of the current state of digital technologies in aquaculture, including vision-based sensors, acoustic-based sensors, and biosensors, for fish tracking, counting, and behaviour analysis. These technologies offer valuable tools for optimizing production efficiency, fish welfare, and resource management in aquaculture. However, each technology has its limitations, such as the sensitivity of vision-based sensors to environmental conditions, the high cost and complexity of acoustic-based sensors, and the potential invasiveness of biosensors.
Despite the advancements in these technologies, significant challenges remain, including the scarcity of comprehensive fish datasets, the lack of unified evaluation standards, and the need for more robust and adaptable systems that can handle the complexities of real-world aquaculture environments. To address these challenges and drive progress in the field, future research should focus on developing diverse and representative datasets, establishing standardized evaluation frameworks, and exploring integrating multiple technologies to create more comprehensive and reliable monitoring systems.
Emerging technologies such as multimodal data fusion, deep learning, and edge computing present exciting opportunities for advancing digital aquaculture. By leveraging these technologies, more accurate, efficient, and practical solutions can be developed for fish tracking, counting, and behaviour analysis, ultimately contributing to the sustainable growth and development of the aquaculture industry. As the field progresses, it is crucial to consider these technologies' ethical implications and environmental impact. Developing solutions that are not only technologically advanced but also sustainable and respectful of animal welfare will be paramount for the future of aquaculture.

\section{AUTHOR CONTRIBUTIONS}
\textbf{Meng Cui}: Conceptualization; writing – original draft. \textbf{Xubo Liu}: Investigation; validation; methodology; data curation. \textbf{Haohe Liu}: Validation; writing – original draft; data curation; formal analysis. \textbf{Jinzheng Zhao}: Validation; writing – review and editing. 
% \textbf{Tao Chen}: Supervision; writing – review and editing; formal analysis. \textbf{Guoping Lian}: Supervision; writing – review and editing.
\textbf{Daoliang Li}: Funding acquisition; validation; project administration. \textbf{Wenwu Wang}: Funding acquisition; validation; project administration; resources; writing-review and editing.

\section{ACKNOWLEDGMENT}
This work was supported by the Research and demonstration of digital cage integrated monitoring system based on underwater robot [China grant 2022YFE0107100], Digital Fishery Cross-Innovative Talent Training Program of the China Scholarship Council (DF-Project) and a Research Scholarship from the China Scholarship Council (202006350248). 

\section{DATA AVAILABILITY STATEMENT}
Since this is a review paper, there is no data available. All information can be found in the cited references.

\section{CONFLICT OF INTEREST STATEMENT}
The authors declare that there are no conflicts of interest.

\begin{bibliography}{mybib}
\bibliographystyle{ieeetr}
\end{bibliography}

\end{document}